\documentclass[showpacs,preprintnumbers,amsmath,amssymb]{revtex4}
\usepackage{graphicx}
\usepackage{dcolumn}
\usepackage{bm}

\begin{document}

\title{ Neutrino generated  dynamical dark energy  with no dark energy field}

\author
{E. I. Guendelman \thanks{guendel@bgu.ac.il} and A.  B. Kaganovich
\thanks{alexk@bgu.ac.il}}
\address{Physics Department, Ben Gurion University of the Negev, Beer
Sheva 84105, Israel}

\date{\today}

\begin{abstract}

Dynamical dark energy (DE) phenomenon emerges as a geometrical effect accompanying the cosmological expansion of nonrelativistic fermionic matter. This occurs without the need for any fluid, like e.g. dynamical scalar field (quintessence, cosmon, etc.), and with conventional form of the Einstein equations   in  contrast to other known geometrical DE models.  The phenomenon results from first principles in the framework of the two measures field theory where,  in the Einstein frame, both fermion masses and the cosmological constant (CC)  turn into functions of the  cold fermion density $n$. This $n$ dependence  becomes negligible  in regular
 (laboratory) conditions but it may have an important role in cosmology.
 In the 4D gravity model where the original action involves only CC  and massive fermions without selfinteraction,  for different (but wide) regions in the parameter space we have found two possible classes of scenarios for the late universe starting from the cold matter domination era. We argue that the fermions which drive the variable CC should be  associated with cold neutrinos disposed in voids and supervoids. The cosmological dynamics of the first class practically coincides with that of the $\Lambda$CDM model, while the dynamics of the second class is of the phantom-like regime with a pseudo-rip scenario. Crossing the phantom divide happens due to a new type of the neutrino DE effect where  neutrinos pass through the state with zero mass and with the vacuum-like EoS $P_{\nu}=-\rho_{\nu}$.

\end{abstract}

\maketitle

\section{Introduction}

Numerous attempts to understand the nature of the discovered cosmic acceleration\cite{acceleration} have originated the common feeling in the cosmology and particle physics community that even though all observational data are in agreement with the simplest model where this phenomena is described by the cosmological constant (CC), the latter is not enough to give adequate answers to \textit{all} challenges of  the  cosmology. Therefore to explain the origin of the acceleration it seems to be impossible to advance without some kind of changes in fundamental of modern physics\cite{Dolgov}. The new physical entity responsible for the present accelerated expansion of the universe was denominated as  the dark energy (DE) and the problem was exposed  intensive study with  vast of models,  for reviews see\cite{{Sahni-Starobinski-review}}-\cite{DEreview_Odintsov}. Using terminology of the review\cite{{Sahni-Starobinski-review}}, to a large extent, all models with variable DE can be  divided into two main classes:
1) physical DE models where DE is associated with some fluid usually realized by means of a new, very weakly interacting physical field.  2) geometrical DE models where effect of DE is realized by means of modified gravity.

In the most of the physical DE models, like e.g. quintessence\cite{quint}, coupled quintessence\cite{amend}, mass varying neutrinos and neutrino DE \cite{MassVar}-\cite{GK_neut_DE}, growing neutrinos\cite{growing-neutr}, chameleon cosmology\cite{chameleon}, phantom\cite{phantom}, quintom\cite{quintom}, the DE entity is modelled by scalar field(s) with very special features (very flat potential, etc.) or/and using  non-scalar systems, e.g. models with specially adapted spinors\cite{spinorDE} . Using of such fields means the exit from the framework of the matter content of the particle Standard  Model.

The geometrical DE models, e.g.  higher dimensional DGP braneworld DE \cite{DGP},\cite{Chimento_Maartens}, $f(R)$ DE \cite{f(R)}, scalar-tensor models\cite{scalar-tensor}, braneworld with Einstein-Gauss-Bonnet gravity\cite{Ein-Gauss_Bonne},   etc. (see also reviews
\cite{Sahni-Starobinski-review},\cite{Modified Gravity and Cosmology},\cite{DEreview_Odintsov}) demonstrate possibilities to realize an effective DE without any field or other matter entity intended to produce DE. However the price for this effect is the emergence of nonconventional 4D gravitational dynamics. Only after certain regrouping of terms having a geometrical meaning to the effective matter source terms, the gravity equations take the Einsteinian GR form.

The model of the present paper is realized in the framework of the two measures field theories (TMT)\cite{GK1}.  We want to stress that  TMT possesses a number of peculiar features which allow to regard it as  a new type of alternative theories. First, the modification of the gravity and matter sectors in TMT is implemented on a common basis. Second,  apart from proceeding in the first order formalism, all the modification in the action of TMT , as a matter of principle,  reduces to the use  of the volume element with two measures: the regular one $\sqrt{-g}$ and the new one $\Phi$ involving additional non-dynamical degrees of freedom (for details see Sec.II).  Third, TMT admits but does not require to involve in the underlying Lagrangian terms and degrees of freedom different from those of the conventional theory (i.e. Einstein gravity and Standard particle model).
Fourth, in the Einstein (physical) frame, the gravity  and all matter field equations have conventional form without the need in regrouping and renaming geometrical and matter source terms. The novelty consists in emergence of variable cosmological constant, very nontrivial form of scalar field potential, fermion selfinteraction and  Yukawa type coupling constants become local function of the fermion density.
Therefore fermions \textit{generically} may possess very unconventional properties that makes it possible, for example, to realize effect of the neutrino driven DE which is the subject of this paper. But  properties of  \textit{fermionic matter in regular conditions  are undistinguished from those in the conventional theory}. The term 'regular conditions' means here that the local fermion energy density $\rho_f$ is tens orders of magnitude larger than the vacuum energy density
$\rho_{vac}$, that is actually a condition which is fulfilled  in all  measurements carried out so far in physics,
including tests of GR. For example in regular conditions,  corrections to fermion mass  have the order $\rho_{vac}/\rho_f$. In this respect, in contrast to other alternative theories, the novelty of the most of the studied so far TMT models is that in order to achieve for undesirable deviations from the conventional theory to be unobservable there is \textit{no  need in special tuning or constraints on the masses and coupling constants in the underlying Lagrangian}.

In our earlier publications  on TMT\cite{G1}-\cite{{emerging}},\cite{GK_neut_DE} we concentrated on studying the scale invariant TMT models.  Here the scale invariance is spontaneously broken by the integration of the equations of motion and we have found that
 the dilaton plays the role of the inflaton field in the early universe and that of the DE field in the late universe. It was shown\cite{GK4} that the Yukawa type effective coupling constant $f$ of the dilaton to matter depends on the local matter density.  For matter in regular conditions the magnitude of $f$ is suppressed by the factor of the order of $\rho_{vac}/\rho_f$. The model yields such  resolution of the fifth force problem automatically, without any special tuning of the parameters.

  In Ref.\cite{GK_neut_DE} devoted to cosmological applications of the model, we have studied the effect of neutrino DE emerging in the course of evolution of the late time universe filled with the homogeneous dilaton field and the cold gas of uniformly distributed non-relativistic neutrinos. A new kind of regime in the cosmological dynamics have been revealed where neutrinos undergo transition to a state with unbounded grows of their mass and this process is accompanied with  reconstruction of the scalar field potential in such a way that this dynamical regime appears to be energetically more preferable than it would be in the case of the universe with no fermions at all.

 Recently another TMT model\cite{G} involving a gravitating scalar field with a Born-Infeld type kinetic term
  and with an arbitrary potential have been constructed.  This model gives a unified picture of dark energy and dark matter reproducing the $\Lambda$CDM picture. A further generalization, including the addition of a tachyon\cite{} allows the formulation of an "inverse quintessence" scenario.

In the present paper we work with the TMT model which may be regarded as a simplified TMT modification of the Standard Model. This is the case because even though the underlying Lagrangian of our simplified model involves only free massive fermions and the CC term, the TMT allows\cite{GK2-1} to extend it by adding all fields and symmetries of Standard Model without changing the main results of the paper. The reason is that the key role in generation of the aforementioned effects of the novelty of TMT (in comparison with conventional theory) belongs to the algebraic constraint which appears due to the new measure of integration $\Phi$. The constraint determines the scalar field $\zeta =\Phi/\sqrt{-g}$ as a local function of $\overline{\Psi}\Psi$ (where $\Psi$ refers to massive fermion fields) and the latter is the only contribution to the constraint from the Standard Model fields. Therefore, at least in the first step of studying the cosmological applications of the model, one can restrict ourself with the simplified model.

In spite of the fact that aside from fermions, our simplified model does not contain any additional dynamical degree of freedom, the variable effective CC $\Lambda_{tot}$ is generated and its local space-time value is determined by the local fermion density. In this paper we explore the cosmological dynamics of the late spatially flat FLRW universe starting from the cold matter domination epoch. $\Lambda_{tot}$ is governed by the density of the cold fermion gas involved in the cosmological expansion. The main goal of this paper is to demonstrate that in the TMT modification of the Standard Model, a variable CC is  generated without the need for additional dynamical degrees of freedom. To our knowledge, this is the first model where this idea is realized, see for example Refs.\cite{Vilenkin},\cite{VCC}.

The organization of the paper is the following. In Sec.II we
present: a review of the basic ideas of TMT; discuss properties of the new degrees of freedom; describe  on qualitative level some details of using the action principle resulting in the equations of motion and the constraint in terms of the original set of variables; explain, again qualitatively, the reasons and the results of transition to the Einstein frame. This allows us, after the formulation of the model in Sec.III, to present the equations of
motion in the Einstein frame at once. Sec.IV is devoted to the very detailed analysis of
physically permitted intervals for the geometrical scalar field $\zeta$. In Sec.V we describe new aspects of the cosmological averaging procedure caused by the appearance of the constraint. Afterwards we present equations of cosmological dynamics resulting from our field model after cosmological averaging and classify possible scenarios of the late universe. In Sec.VI the results of numerical solutions are collected and classified in accordance with qualitative analysis of Sec.V. In Sec.VII we shortly summarize main results of the model that should help  to the reader to see the picture altogether and finally present some preliminary speculations concerning the possibility  to describe the fermion dark matter as a distinctive fermion state in TMT.

\section{Main ideas of the two measures field theory}

\subsection{Two volume measures}

TMT is a generally coordinate invariant theory where the action
may be written in the form\cite{GK1}
\begin{equation}
    S = \int L_{1}\Phi d^{4}x +\int L_{2}\sqrt{-g}d^{4}x
\label{S}
\end{equation}
 including two Lagrangians $ L_{1}$ and $L_{2}$ and two
measures of integration: the usual one $\sqrt{-g}$ and the new one
$\Phi$ independent of the metric. The measure  $\Phi$ being  a scalar density
may be defined as following:
\begin{equation}
\Phi
=\varepsilon^{\mu\nu\alpha\beta}\varepsilon_{abcd}\partial_{\mu}\varphi_{a}
\partial_{\nu}\varphi_{b}\partial_{\alpha}\varphi_{c}
\partial_{\beta}\varphi_{d},
\label{Phi}
\end{equation}
where $\varphi_a(x)$ ($a=1,..,4$) are scalar fields. Since $\Phi$ is a total derivative,
 the shift of $L_{1}$ by a
constant, $L_{1}\rightarrow L_{1}+const.$ has no effect on the
equations of motion. Similar shift of $L_{2}$ would lead to the
change of the constant part of the Lagrangian coupled to the
volume element $\sqrt{-g}d^{4}x $. In  standard GR, this constant
term is the cosmological constant. However in TMT the relation
between the constant
 term of $L_{2}$ and the physical cosmological constant is very non
trivial.

In addition to the above idea concerning the general structure of
the action in TMT, there are only two basic assumptions:
\\
(1) The Lagrangian densities $ L_{1}$ and $L_{2}$ may be functions
of the matter fields,  the metric, the connection (or
spin-connection )
 but not of the "measure fields" $\varphi_{a}$. In such a case, i.e. when the
measure fields $\varphi_{a}$ enter in the theory only via the
measure $\Phi$,
  the action (\ref{S}) has
the infinite dimensional symmetry\cite{GK1}:
$\varphi_{a}\rightarrow\varphi_{a}+f_{a}(L_{1})$, where
$f_{a}(L_{1})$ are arbitrary functions of  $L_{1}$.
\\
(2) We proceed in the first order formalism where all
fields, including metric and connections ( or vierbeins
$e^{\mu}_{a}$ and  spin-connection $\omega_{\mu}^{ab}$ in the
presence of fermions) as well as the measure fields $\varphi_{a}$
are independent dynamical variables. All the relations between
them follow from equations of motion. The independence of the
metric and the connection in the action means that we proceed in
the first order formalism and the relation between connection and
metric is not necessarily according to Riemannian geometry.

Varying the measure fields $\varphi_{a}$, we get
\begin{equation}
B^{\mu}_{a}\partial_{\mu}L_{1}=0 \label{var-phi}
\end{equation}
where
\begin{equation}
B^{\mu}_{a}=\varepsilon^{\mu\nu\alpha\beta}\varepsilon_{abcd}
\partial_{\nu}\varphi_{b}\partial_{\alpha}\varphi_{c}
\partial_{\beta}\varphi_{d}.
\label{Ama}
\end{equation}
Since $Det (B^{\mu}_{a}) = \frac{4^{-4}}{4!}\Phi^{3}$ it follows
that if $\Phi\neq 0$,
\begin{equation}
 L_{1}=sM^{4} =const
\label{varphi}
\end{equation}
where $s=\pm 1$ and $M$ is a constant of integration with the
dimension of mass.

Here we should point out that TMT has some similarity with the recently developed the Lagrange Multiplier Gravity (LMG)\cite{LMG} where there is a Lagrange multiplier field which enforces the condition that a certain Lagrangian is zero. This seems as a particular case of the TMT effect described by Eq.(\ref{varphi}) where the arbitrary integration constant appears.

\subsection{Generic features of TMT}

First of all one should notice
 {\it the very important differences of
TMT from scalar-tensor theories with nonminimal coupling}:
\begin{itemize}
\item In general, the Lagrangian density $L_{1}$ (coupled to the measure
$\Phi$) may contain not only the scalar curvature term (or more
general gravity term) but also all possible matter fields terms.
This means that {\it TMT can modify in general both the
gravitational sector  and the matter sector in the same fashion};

\item If the field $\Phi$ were the
fundamental (non composite) one then the variation of $\Phi$ would
result in the equation $L_{1}=0$ instead of (\ref{varphi}), and
therefore the dimensionful parameter $M$ would not appear.

\end{itemize}

Applying the first order formalism one can show (see for example
Ref.\cite{GK1})  that the resulting relation between metric and
connection, as well as fermion and scalar fields equations involve
 the gradient of the ratio of the two measures
\begin{equation}
\zeta \equiv\frac{\Phi}{\sqrt{-g}} \label{zeta}
\end{equation}
which is a scalar field.
For understanding the structure of TMT it is important to note that TMT (where, as we
supposed, the scalar fields $\varphi_{a}$ enter only via the measure $\Phi$) is a constrained dynamical system.
In fact, the volume measure $\Phi$ depends only upon the first derivatives of $\varphi_{a}$ and this
dependence is linear. The fields $\varphi_{a}$ do not have their own dynamical equations: they are
auxiliary fields. All of their dynamical effect, at least at the
classical level, is displayed only in the following two ways: (a) in generating the constraint; (b) in the appearance of the scalar field $\zeta$ and its
gradient in all equations of motion.

The origin of the constraint is clear enough. There are two
equations containing the scalar curvature: the first one is
Eq.(\ref{varphi}) and the second one follows from gravitational
equations. The constraint is nothing but the consistency condition
of these two equations: eliminating the scalar curvature from these two equations
one obtains $\zeta (x)$ as a local function of
matter fields. The surprising feature of the theory is
 that neither Newton constant nor curvature appear in this constraint
which means that the {\it geometrical scalar field} $\zeta (x)$
{\it is determined by the matter fields configuration}  locally
and straightforward (that is without gravitational interaction).

The appropriate change of the dynamical variables, which includes
a conformal transformation of the metric, results in the formulation of the
theory in a Riemannian (or Riemann-Cartan) space-time. The
corresponding conformal frame we call "the Einstein frame". The
big advantage of TMT is that in the very wide class of models,
{\it the gravity and all matter fields equations of motion take
conventional GR form in the Einstein frame}.
 All the novelty of TMT in the Einstein frame as compared
with the standard GR is revealed only
 in an unusual structure of the cosmological constant, scalar fields
effective potential, masses of fermions  and their interactions
with scalar fields as well as in the unusual structure of fermion
contributions to the energy-momentum tensor: they appear to be
$\zeta$ dependent. This is why the scalar field $\zeta (x)$
determined by the constraint, has a key role in these effects.

\section{The field theory model}

Our simplified TMT model deals with the selfconsistent system involving 4D gravity, the cosmological constant (CC)
 and fermions.
Keeping the general structure of Eq.(\ref{S}) it is convenient to
represent the action $S$ in the following form:
\begin{eqnarray}
&\cdot S&=-\frac{1}{\kappa}\int d^{4}x (\Phi +b\sqrt{-g}) R(\omega ,e)
 -\int d^{4}x \sqrt{-g}\Lambda_0
\nonumber\\
&& +\int d^{4}x (\Phi +k\sqrt{-g})
\frac{i}{2}\sum_{i}\overline{\Psi}_{i}
\left(\gamma^{a}e_{a}^{\mu}\overrightarrow{\nabla}^{(i)}_{\mu}-
\overleftarrow{\nabla}^{(i)}_{\mu}\gamma^{a}e_{a}^{\mu}\right)\Psi_{i}
\nonumber\\
     &&-\int d^{4}x(\Phi +h\sqrt{-g})\sum_{i}\mu_i\overline{\Psi}_i\Psi_i
 \label{totaction}
\end{eqnarray}
where $\Psi_{i}$  are the
fermion fields  of  species labeled by the index $i$;
$\mu_i$ are the mass parameters;
$\overrightarrow{\nabla}=\vec{\partial}+
\frac{1}{2}\omega_{\mu}^{cd}\sigma_{cd}$
(where we ignore gauge fields);  $\kappa =16\pi G$;
 $R(\omega ,V)
=e^{a\mu}e^{b\nu}R_{\mu\nu ab}(\omega)$ is the scalar curvature;
$e_{a}^{\mu}$ and $\omega_{\mu}^{ab}$ are the vierbein and
spin-connection; $g^{\mu\nu}=e^{\mu}_{a}e^{\nu}_{b}\eta^{ab}$ and
$R_{\mu\nu ab}(\omega)=\partial _{mu}\omega_{\nu
ab}+\omega^{c}_{\mu a}\omega_{\nu cb}-(\mu \leftrightarrow\nu)$;
we use the signature $(+---)$;
$\Lambda_0$ is a constant of the dimensionality
$(mass)^4$ which in the standard GR would be the CC.
Notice that there is no need for including similar term with the measure $\Phi$
because such a term is the total derivative and does not contribute into equations of motion.
  Constants $b, k, h$  are non specified
dimensionless real parameters and we will only assume that $b>0$ and they are different but
 have the same order of magnitude
\begin{equation}
b\sim k\sim h.
 \label{sim-parameters}
\end{equation}

 We would like to stress that
except for the modification of the general structure of the action
according to the basic assumptions of TMT,
 {\it we do not introduce into the action}
(\ref{totaction}) {\it any exotic terms and  fields}. Without changing the results of the present paper, one can generalize the model to non-Abelian symmetry
 adding also gauge and Higgs fields
to reproduce the standard model (see \cite{GK2-1}). However, for purposes of the present paper it is enough
to restrict ourself with the simplified model involving only massive fermions and the CC term in the action.
We would like to lay emphasis that refusal from a dynamical scalar field, like quintessence, means that our model does
not exit the matter content of the standard model.

Variation of the measure fields $\varphi_{a}$  yields
Eq.(\ref{varphi}) where $L_{1}$ is now defined, according to
Eq.(\ref{S}), as the part of the integrand of the action
(\ref{totaction}) coupled to the measure $\Phi$.

Equations of motion for fermion fields as well as the equations for the spin-connection
resulting from (\ref{totaction}) in the first order formalism
contain terms proportional to $\partial_{\mu}\zeta$ that makes the
space-time non-Riemannian and equations of motion - non canonical.
However, with the new set of variables
\begin{eqnarray}
&&\tilde{e}_{a\mu}=(\zeta
+b)^{1/2}e_{a\mu}, \quad
\tilde{g}_{\mu\nu}=(\zeta +b)g_{\mu\nu},
\nonumber\\
&&\Psi^{\prime}_{i}=\frac{(\zeta
+k)^{1/2}}{(\zeta +b)^{3/4}}\Psi_{i} ,
\label{ctferm}
\end{eqnarray}
which we call the Einstein frame,
 the spin-connections become those of the
Einstein-Cartan space-time with the metric $\tilde{g}_{\mu\nu}$
The fermion equations also take the standard form of the Dirac equation in
the Einstein-Cartan space-time where now the fermion masses become $\zeta$ dependent
\begin{equation}
m_{i}(\zeta)= \mu_{i}F(\zeta), \qquad where \qquad F(\zeta)=\frac{(\zeta +h)}{(\zeta +k)(\zeta
+b)^{1/2}} \label{m}
\end{equation}
We have supposed here that
\begin{equation}
\zeta +b>0 \qquad \text{and} \qquad \zeta+k>0,
\label{zb-zk-positive}
\end{equation}
 which will be assumed in what follows.

 It is easy to check that  the gravitational equations in the Einstein frame take the standard GR form
\begin{equation}
G_{\mu\nu}(\tilde{g}_{\alpha\beta})=\frac{\kappa}{2}T_{\mu\nu}^{eff}
 \label{gef}
\end{equation}
where  $G_{\mu\nu}(\tilde{g}_{\alpha\beta})$ is the Einstein
tensor in the Riemannian space-time with the metric
$\tilde{g}_{\mu\nu}$; the energy-momentum tensor
$T_{\mu\nu}^{eff}$ is now
\begin{equation}
T_{\mu\nu}^{eff}=T_{\mu\nu}^{(f,can)}+T_{\mu\nu}^{(\Lambda)}
 \label{Tmn}
\end{equation}
where
$T_{\mu\nu}^{(f,can)}$ is the canonical
energy momentum tensor for fermions in curved space-time\cite{Birrell}; $T_{\mu\nu}^{(\Lambda)}$
is the total variable CC term in the presence of fermions
\begin{equation}
T_{\mu\nu}^{(\Lambda)}=\tilde{g}_{\mu\nu}\Lambda_{tot}; \qquad \Lambda_{tot}=\Lambda_{eff}(\zeta)
-\Lambda_{dyn}^{(f)},
\label{Tmn_Lambda}
\end{equation}
where
\begin{equation}
\Lambda_{eff}(\zeta)=
\frac{bsM^{4}
-\Lambda_0}{(\zeta +b)^{2}},
\label{Veff1}
\end{equation}
and
\begin{equation}
\Lambda_{dyn}^{(f)}\equiv  Z(\zeta)F(\zeta)\sum_i \mu_{i}
\overline{\Psi^{\prime}}_i\Psi^{\prime}_i
\label{Lambda-ferm}
\end{equation}
\begin{equation}
Z(\zeta)\equiv \frac{(\zeta -\zeta_1)(\zeta
-\zeta_2)}{2(\zeta +k)(\zeta +h)}
\label{Zeta}
\end{equation}
\begin{equation}
\zeta_{1,2}=\frac{1}{2}\left[k-3h\pm\sqrt{(k-3h)^{2}+
8b(k-h) -4kh}\,\right].
 \label{zeta12}
\end{equation}
Taking into account our basic assumption (\ref{sim-parameters}) concerning the orders of the parameters $b$, $k$ and $h$ we see that
\begin{equation}
\zeta_1\sim\zeta_2\sim b\sim k\sim h.
 \label{sim-parameters_1}
\end{equation}
if no a special fine tuning of the parameters is assumed.

 \textit{The first new effect} we observe here consists in the appearance of \textit{the variable CC} $\Lambda_{eff}(\zeta)$ generated by the integration constant $sM^{4}$ (see Eq.(\ref{varphi}))  and by the $\Lambda_0$ term of the original action (\ref{totaction}).  $\Lambda_{eff}(\zeta)$  is positive provided
 \begin{equation}
bsM^{4}>\Lambda_0.
 \label{bsM_bigger_L0}
\end{equation}

\textit{The second novelty} consists in  generating the
{\em fermion noncanonical} contribution  into the energy
momentum tensor
\begin{equation}
 T_{\mu\nu}^{(f,noncan)}=-\tilde{g}_{\mu\nu}\Lambda_{dyn}^{(f)}
 \label{Tmn-noncan}
\end{equation}
which has the transformation
properties of a CC term but it is proportional
to $\overline{\Psi}^{\prime}_{i}\Psi^{\prime}_{i}$. This is why we refered to it\cite{GK_neut_DE} as
"dynamical fermionic $\Lambda$ term".
The appearance of $T_{\mu\nu}^{(f,noncan)}$ means that {\it even cold fermions generically
possess pressure} $P_f^{(noncan)}$ and
\begin{equation}
P_f^{(noncan)}=-\rho_f^{(noncan)}=\Lambda_{dyn}^{(f)} =\frac{(\zeta -\zeta_1)(\zeta
-\zeta_2)}{2(\zeta +k)^2(\zeta +b)^{1/2}}\sum_i \mu_{i}
\overline{\Psi^{\prime}}_i\Psi^{\prime}_i,
\label{Pfnoncan}
\end{equation}
where $\rho_f^{(noncan)}$ is the noncanonical contribution of fermions into the energy density.

The  scalar field $\zeta$
is determined as a function of
$\overline{\Psi}^{\prime}_{i}\Psi^{\prime}_{i}$  by the following constraint
\begin{equation}
\frac{(b-\zeta)sM^{4}-2\Lambda_0}{(\zeta
+b)^2}=P_f^{(noncan)}.
\label{constraint}
\end{equation}

In the presence of fermions, the r.h.s. of the constraint is nonzero, and therefore $\zeta\neq const$. One can then expect that the variable CC term  $T_{\mu\nu}^{(\Lambda)}$ , Eq.(\ref{Tmn_Lambda}), mimics the dynamical DE effect in the late time universe. This is indeed the case  as we will see later on in the paper. Using the constraint (\ref{constraint}) and Eq.(\ref{Tmn_Lambda}), $T_{\mu\nu}^{(\Lambda)}$ may be represented in the form
\begin{equation}
T_{\mu\nu}^{(\Lambda)}=\tilde{g}_{\mu\nu}\Lambda_{tot}(\zeta), \qquad \Lambda_{tot}(\zeta)=\frac{sM^4\zeta +\Lambda_0}{(\zeta +b)^2}.
\label{T_DE}
\end{equation}
where only the $\zeta$ dependence remains explicitly.

 Let us imagine for the moment that we were succeeded to solve analytically the constraint (\ref{constraint}), which is the 5-th degree algebraic equation with respect to $(\zeta +b)^{1/2}$. Having the function $\zeta=\zeta\left(\sum_i \mu_{i}
\overline{\Psi^{\prime}}_i\Psi^{\prime}_i\right)$  and inserting it to the fermion mass $m_i(\zeta)$, Eq.(\ref{m}), we would obtain a local fermion (very nonlinear) selfinteraction in the Dirac equation. Inserting the same solution for $\zeta$ into $T_{\mu\nu}^{(\Lambda)}$, Eq.(\ref{T_DE}), we would obtain that the DE effect in the late time universe is governed only by the expession $\sum_i \mu_{i}
\overline{\Psi^{\prime}}_i\Psi^{\prime}_i$, while the latter is locally determined by the dynamics of  fermions with the described selfinteraction. In this respect, the key point of the TMT model under consideration consists in \textit{the absence of a DE fluid}. And even though  the form of $\Lambda_{tot}$resembles the DE-like effective  potential \textit{it is not a potential of any dynamical field}.

The described approach runs into the mathematical problem when we want to solve analytically the 5-th degree algebraic equation. Instead, one can proceed with the scalar field $\zeta$ evaluated by means of numerical solving the constraint.   To realize the program of numerical solution, it is useful to notice that in what follows we are going to study the application of this field theory model to the case of {\it nonrelativistic fermions} in the cold fermions (dust) approximation. This means that {\it we neglect the effect of fermion 3-momenta}.  The only component of the canonical fermion energy-momentum tensor $T_{\mu\nu}^{(f,can)}$ which survive in this approximation is the energy density $T_{00}^{(f,can)}=\rho_f^{(can)}$. Making use of the Dirac equation in the same approximation we obtain (see also Eq.(\ref{m}))
\begin{equation}
\rho_f^{(can)}=
 m(\zeta)
\overline{\Psi^{\prime}}\Psi^{\prime}; \qquad  m(\zeta)=\frac{\mu(\zeta +h)}{(\zeta +k)(\zeta
+b)^{1/2}}
\label{rhofcan}
\end{equation}
and  for the sake of simplicity we write down the contribution of a single fermion alone.

To provide the consistent description of the fermion energy density and pressure we have to regard
$\overline{\Psi}^{\prime}\Psi^{\prime}$ as the field operator. Then the standard semiclassical approach to the Einstein eqs.(\ref{gef}) (when the gravity is treated as the classical field) implies the need to evaluate the  matrix element of $\overline{\Psi^{\prime}}\Psi^{\prime}$ between appropriate states of the Fock space resulting in the fermion (particles and antiparticles) number density $n$ and then
 \begin{equation}
\rho_f^{(can)}=
 m(\zeta) n
\label{rhofcan_n}
\end{equation}
 In this respect, the only new element in our model is the appearance of the geometrical scalar field $\zeta$ (treated in semiclassical approach also as the classical field) and the constraint (\ref{constraint}). Proceeding with the constraint in the same way as with the Einstein equation,
 we obtain
\begin{equation}
\frac{(b-\zeta)sM^{4}-2\Lambda_0}{(\zeta
+b)^2}=P_f^{(noncan)}, \quad \text{where} \quad  P_f^{(noncan)}=\Lambda_{dyn}^{(f)}=Z(\zeta)\rho_f^{(can)}
\label{constraint_n}
\end{equation}

 In the case of nonrelativistic fermions, corrections to the cold fermion approximation reduce to the appearance of the Fermi momentum depending factors\cite{FNW-1}, \cite{Goldman} in front of $n$  in Eq.(\ref{rhofcan_n}). For the goals of the present paper it is enough to notice that these factors are positive.

 Variability of the CC term (\ref{T_DE}) is governed by the geometrical, nondynamical scalar field $\zeta(x)$. In its turn, due to the constraint, the local space-time  value $\zeta(x)$ is determined straightforwardly (without gravity) by the local magnitude of the fermion density . And vice versa, the dynamics of the fermion field is  affected (again locally) by $\zeta(x)$ via the $\zeta$ dependent fermion mass, Eq.(\ref{m}).

 This way of solution to the problem does not affect the above conclusion about the absence of a DE fluid but it is rather technical method to get round the pure mathematical problem. In other words, the emergence of $\zeta$ in the numerical solution to the problem does not mean that the model contains an additional dynamical degree of freedom describing DE.

\section{Physically permitted intervals of $\zeta$}

In the TMT  models, the local
magnitude of the fermionic matter density determines the local value of the ratio of two measures
$\zeta=\Phi/\sqrt{-g}$ through the algebraic constraint (\ref{constraint}) (or (\ref{constraint_n})) which does
not involve neither Newton constant nor any other small parameter.
In its turn, the local value of the scalar field $\zeta$ has a
direct impact on the local values of the energy density and
pressure as well as on the fermion mass. For this reason, first of all, we have to be sure
that such  $\zeta$ dependence of observables does not contradicts to experiments. We will make this
in three steps.

\medskip

\textit{Step I}. Let us start from the case of the absence of massive fermions. It follows from
the constraint  that in such a case $\zeta$ is a constant
\begin{equation}
\zeta=\zeta_0 =b-\frac{2\Lambda_0}{sM^4}.
  \label{zete-vacuum}
  \end{equation}
Then the effective cosmological constant in the fermion vacuum reads
\begin{equation}
\rho_{vac} =\Lambda_{eff}(\zeta_0)=
\frac{sM^4}{2(\zeta_0 +b)}=\frac{M^8}{4(sbM^4-\Lambda_0)}.
\label{Veff-vac}
\end{equation}
Since $b>0$ and we also believe $\Lambda_0>0$ then $\rho_{vac}>0$   if $s=+1$ and $bM^4>\Lambda_0$. It means also  $\zeta_0 +b>0$ (see Eq.(\ref{zb-zk-positive})) which provides  that the conformal transformation (\ref{ctferm}) is well defined in the fermion vacuum as well.

 Since the measure $\Phi$ is by definition sign indefinite, the scalar field $\zeta$ may be sign indefinite too.  In some of the TMT models, this effect may be physically important\cite{GK3},\cite{GK6}. But here we restrict ourself\footnote{If there will be a need to expand the parameter region in the future investigation we  can involve the opposite case.} proceeding in the region of the parameter space where $\zeta_0 >0$, that is
 \begin{equation}
 b >\frac{2\Lambda_0}{M^4}.
  \label{M-Lambda}
  \end{equation}
  Then $\zeta_0\sim b$ and Eq.(\ref{sim-parameters_1}) may be completed by adding $\zeta_0$:
  \begin{equation}
\zeta_0\sim\zeta_{1}\sim\zeta_{2}\sim b\sim k\sim h.
 \label{sim-all_parameters_0}
\end{equation}

  A tiny value of the vacuum energy density $\rho_{vac}\sim M^4/4b$ may be achieved  by the choice of a huge value of the dimensionless parameter $b$. For example, if $M\sim 10^3GeV$ then for $\rho_{vac}$ to be of the order of the present day vacuum energy density, the parameter $b$ must be of the order $b\sim 10^{60}$. One should stress that the choice of a huge value of the  parameter $b$ has nothing common with the fine tuning in its usual understanding in field theories (for details see item IV in the Discussion section).

  Using Eqs.(\ref{zete-vacuum}), (\ref{m}) and (\ref{Zeta}) one can rewrite the constraint (\ref{constraint_n}) in the form
 \begin{equation}
\frac{M^4(\zeta_0-\zeta)}{(\zeta
+b)^2}=\Lambda_{dyn}^{(f)}; \qquad \Lambda_{dyn}^{(f)}= P_f^{(noncan)}=Z(\zeta)\cdot\rho_f^{(can)}=Z(\zeta)\frac{\mu(\zeta +h)}{(\zeta +k)(\zeta
+b)^{1/2}}n.
\label{constraint-2}
\end{equation}

  It is very important that due to Eq.(\ref{sim-all_parameters_0}) based on  our basic assumption concerning the parameters of the model, Eq.(\ref{sim-parameters}),  the l.h.s. of the constraint has generically the same order of magnitude as $\rho_{vac}$, Eq.(\ref{Veff-vac}), if no a special fine tuning of the parameters is assumed. But clearly  the l.h.s. of the constraint equals zero in the vacuum.

\medskip

 \textit{Step II}.  Let us now check that the  $\zeta$ dependence of the fermion masses, Eq.(\ref{rhofcan}), has no observable effects in laboratory conditions. To show this, it is enough to notice that in all known so far particle ( as well as nuclear, atomic, etc.) physics measurements we deal with interactions which yield such localization of fermions  that  the local fermion energy density  is many tens orders of magnitude bigger than the vacuum energy density. In what follows, for short, we will refer to such a case simply as \textit{the high fermion density} case. Hence if we want to estimate how the  $\zeta$ dependence of the fermion masses can affect the values of observable mass one should compare $m(\zeta)$ in two cases with different but  high  fermion densities.

In order to provide the validity of the constraint in the case of high fermion density  of a nonrelativistic fermion one should find an appropriate value of
$\zeta$ as the solution of the constraint. With the conclusion that the l.h.s. of the constraint is generically $\lesssim\rho_{vac}$ and with knowledge  of the obtained above expression (\ref{Veff-vac})
for $\rho_{vac}$, it is evident that the structure of the functions involved in the constraint does not allow to solve the problem by choosing $|\zeta|$ to be much bigger or much smaller than $b$.
The only possible way to realize the balance between two sides of the constraint is to allow for $\zeta$ to be very close either to $\zeta_1$ or to $\zeta_2$. Then the factor  $Z(\zeta)$ in the r.h.s. becomes very small and it is able to compensate the large value of $\rho_f^{(can)}$ such that the l.h.s. of the constraint remains $\sim\rho_{vac}$, as we discussed at the end of the step I. So, let us choose $\zeta$ to be very close to $\zeta_1$ and represent it  in the form $\zeta=\zeta_1(1-\delta)$, where $|\delta|\ll 1$. Then the constraint may be written as following:
\begin{equation}
\frac{M^4(\zeta_0-\zeta_1)}{(\zeta_1 +b)^2}=-\frac{\delta\zeta_1(\zeta_1-\zeta_2)}{2(\zeta_1 +k)(\zeta_1 +h)}\rho_f^{(can)}
\label{constraint-3}
\end{equation}

For our purpose it is enough to know only the order of magnitude of $\delta$. On the level of the orders of magnitudes, it follows immediately from Eq.(\ref{constraint-3}) that
\begin{equation}
\delta\sim\rho_{vac}/\rho_f^{(can)}.
 \label{delta-zeta}
\end{equation}
 If we compare two cases of different high  fermion densities $\rho_{f}^{(can)}$ and $\rho_{f}^{(can)\prime}$ with the appropriate values of  $\delta$ and  $\delta^\prime$ then the relative change of the fermion masses is of the order
\begin{equation}
\frac{|\Delta m|}{m}\sim |\delta -\delta^\prime| \lesssim \frac{\rho_{vac}}{\rho_f^{(can)}},
\label{mass-change}
\end{equation}
which is obviously unobservable.

\medskip

\textit{Step III}. After studying two opposite limiting cases of fermion vacuum and high fermion density, we are able to the search of the
physically permitted interval of $\zeta$.
In the case of cold fermions,  it follows from Eqs.(\ref{Tmn}) and (\ref{Tmn_Lambda}) that the total energy-momentum tensor may be written in the perfect fluid form
\begin{equation}
T^{\mu\nu}=(\rho +p) u^{\mu}u^{\nu}-p\tilde{g}^{\mu\nu}=\rho_f^{(can)} u^{\mu}u^{\nu}+\Lambda_{tot}\tilde{g}^{\mu\nu}
\label{perfect-Tmunu_local}
\end{equation}
where $u^{\mu}$ is the four-velocity of the cold fermion and the following expressions for the total
energy density and pressure read :
\begin{equation}
\rho =\rho_f^{(can)}+\Lambda_{tot}(\zeta),
 \label{rho-tot-accept}
\end{equation}
\begin{equation}
p =-\Lambda_{tot}(\zeta).
 \label{p-tot-accept}
\end{equation}
where $\Lambda_{tot}(\zeta)$ is defined by Eq.(\ref{T_DE}).

 The interesting feature of TMT is that by virtue of the constraint (\ref{constraint-2}), {\it the total energy density} as well as {\it the total pressure} in the presence of cold fermions
 may be expressed in the form where they are functions of $\zeta$ alone:
\begin{equation}
\rho =\frac{M^4}{(\zeta +b)^2}\left[\frac{1}{2}(b-\zeta_0)+\zeta+\frac{\zeta_0 -\zeta}{Z(\zeta)}\right]
\label{rho-zeta}
\end{equation}
\begin{equation}
p=-\frac{M^4}{(\zeta +b)^2}\left[\frac{1}{2}(b-\zeta_0)+\zeta\right]
\label{p-zeta}
\end{equation}

Our analysis is based on the following  \textit{natural assumptions}:
 There should be possible transitions from  the high density of cold fermions $\rho_{h.d.}$  (when $\zeta$  is very close to the value $\zeta_1$ or $\zeta_2$)  to the low fermion density ending with asymptotic transition $\rho\rightarrow \rho_{vac}$ as $\zeta\rightarrow \zeta_0$. We suppose that  $\zeta$, as the solution of the constraint (\ref{constraint-2}), is a continuous  function of  $n$ evolving  from $\zeta \approx\zeta_1$ (or $\zeta\approx\zeta_2$) to the regime $\zeta\rightarrow \zeta_0$.  Besides, in the course of the monotonic decay of  $n$, the total energy density $\rho$ must be continuous and positive function of $\zeta$.   This means that $\zeta$ cannot cross over values $\zeta_1$ and $\zeta_2$, where $Z(\zeta)$ equals zero and $\rho$ is singular. As it was already supposed by Eq.(\ref{zb-zk-positive}), $\zeta$ cannot cross also over the values $-b$ and $-k$  because the transformation to the Einstein frame, Eq.(\ref{ctferm}), as well as the fermion mass, Eq.(\ref{rhofcan}), become singular and the constraint turns out to be senseless (it looks then as an equality  of finite and infinite quantities).

 An additional information one can obtain from the constraint (\ref{constraint-2}) taking into account that the fermion number density
 \begin{equation}
n=2\frac{M^4}{\mu}\frac{(\zeta_0-\zeta)(\zeta +k)^2\sqrt{\zeta +b}}{(\zeta +b)^2(\zeta -\zeta_1)(\zeta -\zeta_2)}.
 \label{n_positive}
\end{equation}
is positive. Hence the physically permitted intervals for $\zeta$ must satisfy the condition:
\begin{equation}
(\zeta -\zeta_1)(\zeta -\zeta_2)(\zeta_0 -\zeta)>0.
\label{sign_bigger_zero}
\end{equation}
Notice that by definition $\zeta_2 <\zeta_1$.
Now it is easy to check all possible intervals for $\zeta$, and it turns out that there are only two  regimes for evolution of $\zeta$ which satisfy to our natural  assumptions

\begin{equation}
\text{The case (A):}  \qquad \zeta_2<\zeta_1<\zeta<\zeta_0;
\label{interval_zeta_A}
\end{equation}

\begin{equation}
\text{The case (B):} \qquad \zeta_2<\zeta_0<\zeta<\zeta_1;
\label{interval_zeta_B}
\end{equation}
and the wide region in the parameter space exists where such regimes are possible.

 In the case (A), the noncanonical fermion pressure $P_f^{(noncan)}>0$, while in the case (B) $P_f^{(noncan)}<0$.
 For both cases, in the course of the afore-mentioned monotonic decay of  $n$, $\rho$ changes from $\rho_{h.d.}$ to $\rho_{vac}$ while  $\zeta$ remains of the order of the parameters $b\sim k\sim h$. This means that  the l.h.s. of the constraint (\ref{constraint-2}) has generically the same order (or very close to) as $\rho_{vac}$,
and hence the same is valid for $P_f^{(noncan)}$ and $\rho_f^{(noncan)}$ (see Eq.(\ref{Pfnoncan})):
\begin{equation}
 |\Lambda^{(f)}_{dyn}|=|P_f^{(noncan)}|=|\rho_f^{(noncan)}|\sim \Lambda_{tot}\sim\rho_{vac}
\label{Pf-sim_vac}
\end{equation}
Only in the fermion vacuum $\Lambda^{(f)}_{dyn}=0$ and $\Lambda_{tot}=\rho_{vac}$.
Such the narrow intervals for values of  $\zeta$,  $\Lambda^{(f)}_{dyn}$ and $\Lambda_{tot}$ explain why {\it the noncanonical fermion energy density and pressure are unobservable under regular physics conditions (i.e. in the high fermion density case), but in the cosmology of the late time universe they may be important}.
This result is the consequence of  the constraint (\ref{constraint-2}) where the function $Z(\zeta)$, thanks to its structure, plays the role of a "self-locking retainer" providing  the  narrow intervals for possible values of  the listed quantities.

\section{Cosmological averaging in the late time universe and the cosmological model}

 Let us now return to the end of Sec.III where we discussed  the way to describe the evolution of the local energy density and pressure as governed by the geometrical, nondynamical scalar field $\zeta(x)$. In turn, the latter is evaluated by \textit{numerical} solution of the constraint (\ref{constraint_n}) (rewritten in the form (\ref{constraint-2})) as a function of the local magnitude of the fermion density $n$.
When trying to formulate a cosmological TMT model we have to answer the question
whether the constraint, being very nonlinear in $\zeta$ algebraic equation,
preserves its structure after cosmological averaging, that is it
also has the form as in Eq.(\ref{constraint-2})
\begin{equation}
\frac{M^4(\zeta_0-\bar{\zeta})}{(\bar{\zeta} +b)^2}=\overline{\Lambda_{dyn}^{(f)}}; \qquad \overline{\Lambda_{dyn}^{(f)}}=\bar{P}_f^{(noncan)},
\label{constraint_av}
\end{equation}
where now $\overline{\Lambda_{dyn}^{(f)}}$ and $\bar{P}_f^{(noncan)}$  are  the averaged values of $\Lambda_{dyn}^{(f)}$ and $P_f^{(noncan)}$;
 $\bar{\zeta}$ is the cosmological averaged of its local space-time values $\zeta(x)$ and  now $\bar{\zeta}$  is a function only of the cosmic time.

 Solution to this averaging problem turns out to be simple if we take into account the observed inhomogeneity of the late universe: the existence of regions with  clumped matter and domains where matter is very diluted. For example, in the large scale structure this is manifested in the existence of filaments and voids. The characteristic feature of such a structure is that, on each level of scales, the volume of the low density domains is tens orders of magnitude larger than the volume of regions with clustered matter. On the other hand, due to the self-locking retainer effect (see the end of Sec.IV) the relative differences both in values of $\zeta$ and in values of $\overline{\Lambda_{dyn}^{(f)}}$ in these two types of regions do not exceed one order of magnitude. Hence the cosmological average values   $\bar{\zeta}$ and $\overline{\Lambda_{dyn}^{(f)}}$ with very high accuracy coincide with the values of $\zeta$ and $\Lambda_{dyn}^{(f)}$  in the maximal volume domains of low fermion density. Therefore the averaged constraint indeed has the form of Eq.(\ref{constraint_av}), where $\bar{\zeta}$ and  $\bar{P}_f^{(noncan)}$ practically equal to the appropriate values in the maximal volume domains of low fermion density.

  In the cold matter dominated epoch, in the course of the cosmological expansion, the universe enters the  stage when
 in the maximal volume domains of low fermion density, $\rho_f^{(can)}$ becomes less then $\Lambda_{tot}(\zeta)$. Then the regions of the universe populated with the maximal volume domains of low fermion density start to expand with acceleration which is, properly speaking, \textit{the transition from the cold matter dominated epoch to the DE dominated epoch}. We come to the quite natural conclusion that the \textit{accelerated} cosmological expansion of the universe can proceed due to \textit{accelerated} expansion of voids and other low matter density domains.
 It is evident  that such a manner of the accelerated expansion  becomes more and more dominant in the course of evolution of the late time universe. Some speculations concerning possible approach to resolution of the puzzle why the galaxy formation epoch was about  the  transition to the DE domination epoch are presented in the item VII of the Discussion section. Here we want only to note that if the described mechanism of the accelerated  expansion is right, it means that the transition to it could happen only after the voids are formed. But the latter must be accompanied by simultaneous formation of clustered matter.

 In the present paper we do not address the averaging problem of Einstein equations\cite{backreaction}. Since the Einstein Eqs.(\ref{gef}) has the standard GR form,  the first Friedmann equation in the spatially flat FLRW universe is as usual
\begin{equation}
\left(\frac{\dot{a}}{a}\right)^{2}=\frac{1}{3M_{p}^{2}}\bar{\rho}.
 \label{FRW}
\end{equation}
where $\bar{\rho}$ should be the averaged total energy density.

 We would like to apply our model to the scenario which starts from the cold matter (dust) domination era, that is when  the (dark) matter energy density is larger than the DE density. Let us suppose that the dark matter consists of fermions. Regarding the galaxy dark mater as the cold one we assume that its local conventional (thermal) pressure  is negligible with respect to its local energy density. Then the local noncanonical energy density and pressure of the clumped fermionic dark matter being of the order of $\rho_{vac}$ (see Eq.(\ref{Pf-sim_vac})) are certainly negligible as well. It was shown in the step II of Section III that the $\zeta$ dependence of fermion masses is negligible in space regions where fermion energy density is much larger than the vacuum energy density,  and values of $\zeta$ are there very close to $\zeta_1$ or $\zeta_2$. Galaxies and even filaments can be approximately regarded as such regions. Then the cosmological averaged energy density of the clustered matter scales as $1/a^3$ and we will describe it in the form
\begin{equation}
\bar{\rho}_{cl}=\frac{M_{cl}^4}{a^3}
\label{rho-clust}
\end{equation}
where $M_{cl}^4$ is the phenomenological parameter \footnote{Notice that $M_{cl}$ has the dimensionality of mass and it is the only phenomenological parameter of our model. All the others are the Lagrangian's  parameters and besides there is the integration constant $M$.}  which allows to fit the duration of the cold matter domination era.

Averaging of the second equation in (\ref{constraint-2}) yields
 \begin{equation}
 \overline{\Lambda_{dyn}^{(f)}}=Z(\bar{\zeta})\bar{\rho}_f^{(can)},
 \label{Lambda-canon-av}
\end{equation}
where the cosmological averaged of the canonical energy density $\bar{\rho}_f^{(can)}$ of the cold fermion matter disposed in the  maximal volume domains of low fermion density is obtained by averaging Eq.(\ref{rhofcan_n}):
\begin{equation}
\bar{\rho}_f^{(can)}=m(\bar{\zeta})\bar{n}=\frac{(\bar{\zeta} +h)}{(\bar{\zeta} +k)(\bar{\zeta}
+b)^{1/2}}\mu \bar{n},
 \label{rho-canon-av}
\end{equation}
where $\bar{n}\sim 1/a^3$. Therefore $\bar{\rho}_f^{(can)}$ scales in a way different from $1/a^3$.

 The local behavior of the total variable CC term  in the presence of fermions, $T_{\mu\nu}^{(\Lambda)}(x)$, Eqs.(\ref{Tmn_Lambda}) and (\ref{T_DE}), which as we hope is able to mimic the effect of dynamical DE, is governed by $\zeta(x)$. In the  FLRW universe, the averaged components of $T_{\mu\nu}^{(\Lambda)}$ are obviously governed by $\bar{\zeta}$ and they read
 \begin{equation}
\bar{p}_{(\Lambda)}=-\bar{\rho}_{(\Lambda)}=-\Lambda_{tot}(\bar{\zeta}),
\label{p_rho_av}
\end{equation}
where
\begin{equation}
\Lambda_{tot}(\bar{\zeta})=\Lambda_{eff}(\bar{\zeta})
-\overline{\Lambda_{dyn}^{(f)}}=\frac{M^4}{(\bar{\zeta} +b)^2}\left[\frac{1}{2}(b-\zeta_0)+\bar{\zeta}\right],
\label{Leff_av}
\end{equation}
and $\Lambda_{eff}$ is determined by Eq.(\ref{Veff1}).

 We would like to stress here that solving the constraint (\ref{constraint_av}) for $\bar{\zeta}$ we obtain it as a function of $\bar{n}$: $\bar{\zeta}=\bar{\zeta}(\bar{n})$. We come to \textit{the crucial role result of the present paper}:   $\Lambda_{tot}$ is a function of the number density $\bar{n}$ of the cold fermion matter disposed in the  maximal volume domains  of low fermion density:
\begin{equation}
\Lambda_{tot}=\Lambda_{tot}(\bar{n})
\label{Ltot-func-n}
\end{equation}
and  \textit{the dynamical DE effect is driven by fermion degrees of freedom}.

Ignoring the averaged value of the clustered matter pressure, we have that the total averaged pressure
\begin{equation}
\bar{p}=\bar{p}_{(\Lambda)}=-\Lambda_{tot}.
\label{p-av}
\end{equation}

The total canonical matter energy density consists of contributions of the clustered (dark) matter and the canonical energy density $\bar{\rho}_f^{(can)}$ of the cold fermions  disposed in the  maximal volume domains:
\begin{equation}
\bar{\rho}_m=\bar{\rho}_{cl}+ \bar{\rho}_f^{(can)}
\label{rho-tot-normal-av}
\end{equation}

The total averaged energy density, which in the flat universe is the critical one, is then
\begin{equation}
\bar{\rho} =\bar{\rho}_m+\Lambda_{tot}(\bar{n}),
\label{rho-av}
\end{equation}
and fractions of the normal matter  energy density and  the DE density are then defined as usual:
\begin{equation}
\Omega_m =\frac{\bar{\rho}_m}{\bar{\rho}} \qquad \text{and} \qquad \Omega_{DE} =\frac{\Lambda_{tot}(\bar{n})}{\bar{\rho}}
\label{Omega-definitions}
\end{equation}

Making use of the constraint, Eq.(\ref{constraint_av}), one can rewrite $\bar{\rho}$ and $\bar{p}$ in the form where  the contributions from the  maximal volume domains  depend only on the averaged scalar field $\bar{\zeta}$ (cf. Eqs.(\ref{rho-zeta}), (\ref{p-zeta})):
\begin{equation}
\bar{\rho} =\bar{\rho}_{cl}+\frac{M^4}{(\bar{\zeta} +b)^2}\left[\frac{1}{2}(b-\zeta_0)+\bar{\zeta}+\frac{\zeta_0 -\bar{\zeta}}{Z(\bar{\zeta})}\right]
\label{rho-zeta-av}
\end{equation}
\begin{equation}
\bar{p}=-\frac{M^4}{(\bar{\zeta} +b)^2}\left[\frac{1}{2}(b-\zeta_0)+\bar{\zeta}\right]
\label{p-zeta-av}
\end{equation}

The constraint (\ref{constraint_av}) is the  5-th degree algebraic equation   which determines $\zeta$ as an implicit function of $\bar{n}$. Since the constraint is valid for any value of $t$,  the time derivatives of both sides of the constraint are also equal at all t.  By virtue of this evident fact and using the Friedmann equation (\ref{FRW}) we obtain the first order differential equation for $\bar{\zeta}$
\begin{equation}
B\dot{\bar{\zeta}}=\frac{1}{M_p}\left(\zeta_0-\bar{\zeta}\right)\sqrt{3\bar{\rho}}
\label{diff-eq-zeta}
\end{equation}
where
\begin{equation}
B=1+\left(\zeta_0-\bar{\zeta}\right)\left(\frac{1}{\bar{\zeta} -\zeta_1}+\frac{1}{\bar{\zeta} -\zeta_2}+\frac{3}{2(\bar{\zeta} +b)}-\frac{2}{\bar{\zeta}+k}\right)
\label{B}
\end{equation}
It is much more convenient to work with the  differential equation  than to proceed with the constraint itself.

 Notice that
 \begin{equation}
 \bar{\rho}+ \bar{p}= \bar{\rho}_m,
 \label{p+rho}
\end{equation}
where $\bar{\rho}_m$ is determined by Eq.(\ref{rho-tot-normal-av}). Then the total averaged energy-momentum tensor may be written in the perfect fluid form
\begin{equation}
\bar{T}^{\mu\nu}=\bar{\rho}_m u^{\mu}u^{\nu}+\Lambda_{tot}\bar{\tilde{g}}^{\mu\nu}
\label{perfect-Tmunu}
\end{equation}
where $\bar{\tilde{g}}_{\mu\nu}=diag(1,-a^2,-a^2,-a^2)$, the four-velocity $u^{\mu}$ of the cold fermion gas in the co-moving frame is $u^{\mu}=(1,0,0,0)$ and $\Lambda_{tot}$ is determined by Eq.(\ref{Leff_av}).
The structure of the cosmological model under consideration is very unusual. For example,  the fermion masses and  $\Lambda_{tot}$ are $\zeta$ dependent and evolve in the course of cosmological expansion.  In spite of this, the standard GR equation of the energy-momentum conservation
 \begin{equation}
 \dot{\bar{\rho}}=-3\frac{\dot{a}}{a}(\bar{\rho} +\bar{p})
 \label{en-mom-conserv}
\end{equation}
is satisfied. This statement may be checked by means of bulky enough calculations making use
   $\bar{\rho}$ and $\bar{p}$, Eqs.(\ref{rho-zeta-av}), (\ref{p-zeta-av}), the constraint (\ref{constraint_av}) and Eqs.(\ref{diff-eq-zeta}) and (\ref{FRW}).

   Suppose for the moment that we have solved the system of Eqs.(\ref{FRW}) and (\ref{diff-eq-zeta}). Then we have $\bar{\rho}$ and $\bar{p}$ as functions of $\bar{\zeta}$ which in turn is a function of cosmic time (or of the scale factor). These functions may be treated as the function $\bar{p}=\bar{p}(\bar{\rho})$ given in the parametric form. We would like here to stress  again that even though we deal effectively with a kind of barotropic fluid, the model does not involve any  degree of freedom intended to be dynamical DE.

To solve the system of Eqs.(\ref{FRW}) and (\ref{diff-eq-zeta}) we need an additional information which allows to choose the initial condition for $\zeta$.   This information one can obtain from the constraint, Eq.(\ref{constraint_av}),  taking into account that at the very beginning of the scenario described by our model, i.e. in the cold fermionic matter domination era,  $m(\bar{\zeta})\bar{n}$ is many orders of magnitude larger than the l.h.s. of the constraint (\ref{constraint_av}) which  is of the order of (or very close to) the present vacuum energy density (see Eq.(\ref{Pf-sim_vac})). This is possible if $Z(\bar{\zeta})$, Eq.(\ref{Zeta}), is very close to zero in  the cold matter domination era. Taking into account the physically permitted intervals of $\zeta$, Eqs.(\ref{interval_zeta_A}) and (\ref{interval_zeta_B}),  we must choose the initial value $\zeta_{in}$ to be  very close to $\zeta_1$. Then the EoS $w=\bar{p}/\bar{\rho}$ is very close to zero  and negative (see Eqs.(\ref{rho-zeta-av}), (\ref{p-zeta-av}) and (\ref{Zeta})). In the opposite limiting case, when $a(t)\rightarrow \infty$, it follows from the constraint that $\bar{\zeta}\rightarrow \zeta_0$, where $\zeta_0$ is the vacuum value of  $\bar{\zeta}$ defined by Eq.(\ref{zete-vacuum}). One can immediately see from Eqs.(\ref{rho-zeta-av}) and (\ref{p-zeta-av}) that $w$   asymptotically approaches $w=-1$.

Using the results of the qualitative analysis in Sec.IV (Step III) one can classify  the dynamics of the late time universe in the correspondence with the region of the parameter space:

\begin{itemize}

\item
 \textit{The I.A type of scenario with $w>-1$}, when $-b<\zeta_1$,\,  $-k<\zeta_1$ and  $-h<\zeta_1<\zeta_0$ and $\bar{\zeta}$ satisfies the inequalities (A), Eq.(\ref{interval_zeta_A}),
 \begin{equation}
 \zeta_2<\zeta_1<\bar{\zeta}<\zeta_0;
\label{IA}
\end{equation}

\item
 \textit{The I.B type of scenario with $w>-1$}, when $-b<\zeta_0$,\, $-k<\zeta_0$ and  $-h<\zeta_0<\zeta_1$ and $\bar{\zeta}$ satisfies the inequalities (B), Eq.(\ref{interval_zeta_B}),
 \begin{equation}
 \zeta_2<\zeta_0<\bar{\zeta}<\zeta_1;
\label{IB}
\end{equation}

  The I.A and I.B types of scenarios are very similar:
the total energy density
  $\bar{\rho}$ monotonically decreases to $\rho_{vac}$, Eq.(\ref{Veff-vac}), and
 the EoS monotonically decreases to $w=-1$. The difference consists in the opposite behavior of $\bar{\zeta}$: in the I.A type of scenario $\bar{\zeta}$ increases approaching the vacuum value $\zeta_0$, while in the I.B type of scenario $\bar{\zeta}$ decreases approaching  $\zeta_0$. Such behavior is dictated by Eq.(\ref{diff-eq-zeta}): in the I.A case $\dot{\bar{\zeta}}>0$, while in the I.B case $\dot{\bar{\zeta}}<0$.

\item
\textit{The II, Phantom-like type of scenario}, when  $-b<\zeta_0$,\, $-k<\zeta_0$, \, $\zeta_0<-h<\zeta_1$ and $\bar{\zeta}$ \, - \, as in inequalities (\ref{IB}). It follows from Eq.(\ref{diff-eq-zeta}) that $\dot{\bar{\zeta}}<0$, i.e. $\bar{\zeta}$ decreases approaching  $\zeta_0$.
  It is easy to see from Eqs.(\ref{rho-canon-av}), (\ref{constraint_av}),  (\ref{Lambda-canon-av}) and  (\ref{Zeta}) that
 \begin{equation}
 sign\left(m(\bar{\zeta})\bar{n}\right)= sign\left[\frac{1}{Z(\bar{\zeta})}(\zeta_0-\bar{\zeta})\right]=sign(\bar{\zeta}+h)
 \label{sign-zeta-h}
\end{equation}
In the course of decreasing, $\bar{\zeta}$ crosses the value $-h$.
At this moment, both the effective fermion mass  and $\bar{\rho}_f^{(can)}$, Eq.(\ref{rho-canon-av}), cross zero values and become negative, but $\bar{\rho}_f^{(can)}+\bar{\rho}_{(\Lambda)}$ remains positive provided that the parameters are chosen such that, in addition to the listed conditions,
\begin{equation}
\frac{1}{2}(b-\zeta_0)+\bar{\zeta}+\frac{\zeta_0 -\bar{\zeta}}{Z(\bar{\zeta})}>0
\label{rho_positive}
\end{equation}
not only for $-h<\bar{\zeta}<\zeta_1$ but also when $\zeta_0<\bar{\zeta}<-h$. By means of numerical estimations one can check that there is a wide range of parameters where all the above conditions are satisfied.

Taking into account that $\bar{p}_{(\Lambda)}+\bar{\rho}_{(\Lambda)}=0$ we have for the {\it total} EoS $w=p/\rho$
\begin{equation}
w+1=\frac{1}{\bar{\rho}}[\bar{\rho}_{cl}+\bar{\rho}_f^{(can)}]
\label{EoS}
\end{equation}
At the moment when $\bar{\zeta}=-h$,  the fermions  possess only the noncanonical  pressure and energy density and their EoS coincides with the cosmological constant EoS. Such a vacuum-like fermion state has nothing common with usual conception of what is a particle. One can suppose that from all known particles, only neutrinos in  maximal volume domains of low fermion density  are able to reach such exotic state. This state may be treated as the limiting realization of the idea of the neutrino dark energy\cite{Neutr-dark}-\cite{GK_neut_DE}.
If the cosmological averaged energy density of the clumped matter $\bar{\rho}_{cl}$ is less than  the DE density $\bar{\rho}_{(\Lambda)}$ (that is in the agreement with the present day cosmological data) then soon after  $\bar{\zeta}$ crosses the value $-h$,  $w$ should cross the value $w=-1$. Further cosmological expansion proceeds with $|m(\bar{\zeta})\bar{n}|\rightarrow 0$ (that causes $\bar{\zeta}-\zeta_0\rightarrow 0$) and $\bar{\rho}_{cl}\rightarrow 0$. Therefore $\bar{\rho}$ should pass through minimum that is followed by monotonic increasing with $\bar{\rho}\rightarrow\rho_{vac}$ from below  and  $w\rightarrow -1$ from below  as well.
Notice that here $\rho_{vac}$ is determined by  Eq.(\ref{Veff-vac}) and, as we have seen in Sec.IV, Eq.(\ref{Pf-sim_vac}), it is of the order of (or very close to) the present vacuum energy density.
 Such an asymptotic behavior belongs to the pseudo-rip type of the future of the universe, according to the classification of Refs.\cite{Frampton},\cite{Review_Phantom_scenarios}.
    Such a phantom-like cosmology of the very late time universe is realized here without phantom scalar field\cite{phantom} as well as without modification of the Friedmann equation\cite{Chimento_Maartens} based on the braneworld model\cite{DGP}.
    However since the cosmological dynamics of the model is  governed by the effective barotropic fluid (see the discussion after Eq.(\ref{en-mom-conserv})) all problems of singularity and instabilities (with respect to cosmological perturbations) related to the possibility of crossing the phantom divide\cite{Vikman},\cite{CPhantomDivide} are relevant here and  have to be studied. This will be done in a separate publication.

\item
\textit{The III, Phantom-like scenario with unlimited grows of $|m(\bar{\zeta})|$} is realized with the same conditions as in the II type of scenarios except for the very special fine tuned choice  $\zeta_0 =-k$. In this case, the scenario is similar to the II type of scenarios but one can see that in the last stage, as $a\rightarrow\infty$ \, and $\bar{\zeta}\rightarrow\zeta_0 =-k$, the neutrino mass is negative and behaves as $m(\bar{\zeta})\sim (\bar{\zeta}+k)^{-1}$; \, the fermion number density $n$, Eq.(\ref{n_positive}), behaves as $n\sim 1/a^3\sim (\bar{\zeta}+k)^3$. Therefore the canonical neutrino energy density, Eq.(\ref{rho-canon-av}), being negative approaches zero as $\bar{\rho}_f^{(can)}\sim (\bar{\zeta}+k)^2\sim 1/a^2$, while $\rho_{(\Lambda)}\rightarrow\rho_{vac}$. Notice that this qualitative analysis is consistent with Eq.(\ref{diff-eq-zeta}) since in the regime when $\bar{\zeta}$ being larger than $-k$ appears to be also close to $-k$ then  the equation yields
$\dot{\bar{\zeta}}\rightarrow 0^{-}$, i.e. $\bar{\zeta}$ is forced to tend to $-k$.

\end{itemize}

\section{The model results computed numerically}

The results of numerical computations  are presented in Figs.1-17 where we have used the dimensionless units obtained by the following changes:
 \begin{equation}
 \frac{\zeta}{b}\rightarrow\zeta; \quad \frac{k}{b}\rightarrow k; \quad  \frac{h}{b}\rightarrow h; \quad
 \frac{\Lambda_0}{bM^4}\rightarrow\Lambda_0; \quad \frac{b}{M^4}M_{cl}^4\rightarrow M_{cl}^4;
 \nonumber
 \end{equation}
 \begin{equation}
 \text{all others energy densities and pressures:} \quad \frac{b}{M^4}\rho\rightarrow\rho \quad \text{and} \quad
 \frac{b}{M^4}p\rightarrow p.
 \label{units}
\end{equation}

\begin{itemize}

\item
 \textit{The  typical features of the I.A type of the scenario} are illustrated in Figs. 1-4 with the choice of the parameters $\Lambda_0 =-1.5$, $k=2$, $h=0.3$,
 $M_{cl}^4=10^{-5}$. Then $\zeta_1=2.3114$, $\zeta_2 =-1.2114$. The initial value of $\bar{\zeta}$ is chosen
to be $\bar{\zeta}_{in}=2.312$.

One can see that for $\ln a\gtrsim -4$, which corresponds to $-1\leq w\lesssim -0.05$, \, $\zeta\approx const. =\zeta_0$ where $\zeta_0$ is the vacuum value of $\zeta$ (see Eq.(\ref{zete-vacuum})). Therefore in the whole interval $-1\leq w\lesssim -0.05$ \, both $\Lambda^{(f)}_{dyn}$ and $\Lambda_{tot}$ remain practically constant and with high accuracy $\Lambda_{tot}\thickapprox \rho_{vac}$ where $\rho_{vac}$ is defined by Eq.(\ref{Veff-vac}. It means that this type of scenario practically coincides with the appropriate $\Lambda$CDM model with the same value of the CC defined by Eq.(\ref{Veff-vac}).

\item
\textit{The  typical features of the I.B type  of the cosmological evolution scenario} are illustrated in Figs. 5-10 with the choice of the parameters: $\Lambda_0=-0.2$; $k=-0.3$; $h=-1$; $M_{cl}^4=5\cdot 10^{-5}$; Then $\zeta_1=3.0595$; $\zeta_2=-0.3595$. The initial value of $\bar{\zeta}$ is chosen
to be $\bar{\zeta}_{in}=3.0594$.

One can see that $\zeta$ is in a constancy regime  $\zeta\approx const. =\zeta_0$ for $\ln a\gtrsim -3$. The latter corresponds to $w\lesssim -1/3$, that is in the whole DE domination era with high accuracy $\Lambda_{tot}\thickapprox \rho_{vac}=const.$ and besides $\Lambda^{(f)}_{dyn}\thickapprox 0$. Therefore this type of scenario also  practically coincides with the appropriate $\Lambda$CDM model with the value of the CC equal to $\rho_{vac}$. To show explicitly that in the parameter region used in the I.B type of scenario, the cosmological dynamics practically coincides with the $\Lambda$CDM model, we present in the same  figures  5-8 and 10 solutions of our TMT (solid lines) and $\Lambda$CDM (dot-dashed lines) together.  In order to avoid superimposition of the graphs of the TMT and $\Lambda$CDM models, the initial values of the cold (dark) matter energy density in $\Lambda$CDM model have been chosen to be slightly bigger than those in the appropriate scenario of our TMT model.

\item{The  typical features of the II, phantom-like type of the cosmological evolution scenario}  are illustrated in Figs. 11-16 with the choice of the parameters: $\Lambda_0=-0.2$; $k=-0.3$; $h=-6$; $M_{cl}^4=5\cdot 10^{-8}$; Then $\zeta_1=18.2267$; $\zeta_2=-0.5267$. The initial value of $\bar{\zeta}$ is chosen to be $\bar{\zeta}_{in}=18.2266$.

    $\zeta$ is in a constancy regime  $\zeta\approx const. =\zeta_0$ for $\ln a\gtrsim -3$. The latter corresponds here to $w\thickapprox -1$.   In the course of the evolution in the interval $-7\lesssim \ln a\lesssim -3$, \, $\zeta$ decays from $\zeta\thickapprox 18$ to $\zeta\thickapprox \zeta_0 =1.4$. The appropriate monotonic growth of $\Lambda_{tot}$ is from $\Lambda_{tot}\thickapprox 0.05$ to $\Lambda_{tot}\thickapprox \rho_{vac}\thickapprox 0.21$, while $\Lambda^{(f)}_{dyn}$ changes non-monotonically (see Fig. 13). However the crossing of phantom divide happens due the sign change of mass of  neutrinos disposed in the maximal volume domains of low density (Fig. 15), as it was discussed after Eq.(\ref{EoS}). After passing their minima, $w$ and $\rho$ increase asymptotically with cosmic time approaching from below $w=-1$ and  $\rho=\rho_{vac}$ respectively and thus providing a pseudo-rip scenario\cite{Frampton}.

    To show explicitly the difference of  the II type of scenario with respect to  the $\Lambda$CDM model, we present in the same  figures  11-14 and 16 solutions for the appropriate $\Lambda$CDM model (dot-dashed lines).

\item{The  typical features of the III, phantom-like type of the cosmological evolution scenario}  are similar to those of the II type. The only difference consists in the fine tuning of the parameters such that $\zeta_0 =-k$ which yields the unlimited approach to $-\infty$  of the neutrino mass in the maximal volume domains of low density. This effect is  illustrated in Fig.17.

\end{itemize}

The results of the numerical solutions confirm the classification  and properties of possible cosmological scenarios obtained by the qualitative analysis in end of Sec.V.
The key point is {\it the confirmation of the self-locking retainer effect}: as the total energy density decays from a very large value  at the cold matter dominated epoch up to the very low value at the DE dominated epoch, the change of the averaged scalar $\bar{\zeta}$ does not exceed one order of magnitude. As the direct consequence of this effect, the changes of the averaged values of both the dynamical fermionic $\Lambda$ term, $\Lambda^{(f)}_{dyn}$, and the total variable CC, $\Lambda_{tot}$, during the studied  history of the universe does not exceed one order of magnitude too. Remind that in our model the effect of the dynamical DE described by $\Lambda_{tot}$ is  driven by the cold fermions whose density, in turn, changes from the value at the cold matter dominated epoch up to the density of cold neutrinos in voids which tends to zero with the cosmological expansion. Therefore the {\it total} EoS $w=p/\rho$ evolves mainly due to  the evolution of the dark matter energy density while the effect of the change in the dynamical DE  is minor. This is in an agreement with observational constraints on the DE dynamics studied in Ref.\cite{Tsujikawa} where it has been shown that models with a fast varying DE equation-of-state are not favored over the $\Lambda$CDM model.

\section{Summarizing the results  and discussion}

\textit{I. Fermions in TMT.}  One of the peculiarities of the TMT consists in the perfectly novel features of massive fermions, and the geometrical scalar field  $\zeta$, Eq.(\ref{zeta}), plays the main role in this respect:

 1. The fermion mass is $\zeta$ dependent, while the local value  $\zeta(x)$ is determined via the constraint as a function of the local fermion density. As the local fermion energy density $\rho_f$ is many orders of magnitude larger than the vacuum energy density $\rho_{vac}$, the fermion mass is constant up to corrections of the order of $\rho_{vac}/\rho_f$, and the local value  $\zeta(x)$ must be very close either to $\zeta_1$ or to  $\zeta_2$ (see the step II of Sec.IV). Thus, such "high density" fermion states belong to the regular particle physics situation.

 2. As $\rho_f$ becomes comparable with $\rho_{vac}$, the fermion mass becomes $\zeta$ dependent, and due to the constraint this means that the fermion mass depends upon the fermion density. For short, we called such kind of fermion state as the low density one.

 It follows from the last two items that in order to describe the fermion state in the quantum field theory formulated in the framework of TMT, in addition to the usual quantum numbers one should add the value of $\zeta$.

 3. In addition to the canonical energy-momentum tensor, fermions possess also the dynamical fermionic $\Lambda^{(f)}_{dyn}$ term,  Eq.(\ref{Tmn-noncan}), and therefore the appropriate noncanonical pressure $P_{(f,noncan)}=\Lambda^{(f)}_{dyn}$ and the noncanonical energy density $\rho_{(f,noncan)}$ satisfy  the equation  $P_{(f,noncan)}=-\rho_{(f,noncan)}$. We have shown that  $\rho_{(f,noncan)}$ and $\rho_{vac}$ are \textit{always} of the same or very close orders of magnitude. For that reason the effect of the noncanonical energy-momentum tensor is unobservable in all particle physics measurements till now but it has a crucial role in the cosmology of the late time universe.

\textit{II. Dynamical DE imitated by the geometrical effect of TMT in the presence of massive fermions.}
 The matter content of the 4D TMT model studied in the present paper is reduced to massive fermions though radiation could be added without altering the results because gauge fields do not contribute to the constraint. The important feature of the model is that it does not involve any fluid intended to describe the DE. In the case of the absence of fermions, it follows from the constraint that $\zeta =const$ and then the model contains the constant CC. However, in the presence of fermions, on account of the constraint, $\zeta$ becomes a function of the local fermion density. In its turn, the constant CC turns into the  $\zeta$ dependent CC, i.e. on account of the constraint, the CC becomes dependent of the fermion density. This CC together with $\Lambda^{(f)}_{dyn}$ form the effective  variable CC, $\Lambda_{tot}$, Eq.(\ref{T_DE}) depending upon the fermion density. Note that since  $\Lambda_{tot}$ contains the term $\Lambda^{(f)}_{dyn}$ it would be incorrect, as usually, to treat $\Lambda_{tot}$ as the variable {\it vacuum} energy density. It seems to be more natural to regard $\Lambda_{tot}$ as the effective dynamical DE imitated by the geometrical effect of TMT in the presence of massive fermions and driven by the fermion density.

\textit{III. Cosmological averaging and cosmology of the late universe.}

 1. We have concentrated on the cosmological outputs of the key effect of the  model: generating of the variable CC $\Lambda_{tot}$. We have shown that the constraint keeps its form after the cosmological averaging and in the course of the cosmological expansion, the cosmological average of $\Lambda_{tot}$ with the very high accuracy may be regarded as a function of the  density $n$ of cold fermions disposed in the maximal volume domains with the low fermion density. We have proposed to associate these fermions with neutrinos because it seems to be natural to assume that, in the course of cosmological expansion, from all known fermions only nonrelativistic neutrinos are able to survive without interactions up to the state when their canonical (thermal) pressure is negligible in comparison with their noncanonical pressure $\Lambda^{(f)}_{dyn}$ . If this assumption is true then the dynamical DE effect is driven by neutrinos disposed in voids and supervoids.

We have found the  permitted by TMT intervals of $\zeta$ where cold fermions can evolve continuously from  the high density  to the low fermion density state ended with asymptotic transition $\rho\rightarrow \rho_{vac}$. According to these intervals of $\zeta$ we have classified possible scenarios of the late universe.
Almost in all types of scenarios (except for the I.A type), $P_{(f,noncan)}=\Lambda^{(f)}_{dyn}$ is negative which means that $\Lambda^{(f)}_{dyn}$ describes {\it the neutrino DE} and the latter is the essential fraction of the total effect.

2. For the I.B type of scenario, where the {\it total} EoS $w=p/\rho$ decays \textit{monotonically} from $w\approx 0$ to $w= -1$, there exists a wide range in the parameter space such that the differences in the behavior of the total EoS and of the total averaged energy density from those  in the $\Lambda$CDM model are practically unobservable.

3.   In another wide region of the parameter space, the model admits  a phantom-like cosmological scenario
where the crossing of the phantom divide is caused by changing the sign  of the effective neutrino mass in the course of the cosmological expansion.
 As the result, at this moment the cold neutrinos energy density and pressure satisfy exactly the vacuum EoS that means disappearance of neutrinos as particles (excitations) in a sense this term has in the quantum field theory. Such a limiting realization of the idea of the neutrino DE \cite{Neutr-dark}-\cite{GK_neut_DE} deserves more detailed investigation. Even though the sign of fermion mass in the relativistic quantum mechanics is a question of choice, it remains unclear the physical meaning of  the new cosmological dynamics effect we observe in our model:  "degeneration" of neutrino with positive effective mass with subsequent  "regeneration" with negative effective mass. Nevertheless this strange effect, from our viewpoint, seems to be less dangerous than the use of the phantom scalar field\footnote{ Notice that the total measure $\Phi +k\sqrt{-g}$  in the fermion kinetic term of the underlying action (\ref{totaction}) does not change the sign at the moment of crossing the phantom divide. In this point the model of the present paper is drastically different from the model studied in Refs.(\cite{GK3}), (\cite{GK6}). There we have studied the scale invariant TMT model involving dynamical scalar field (dilaton) which plays the role of the inflaton in the early universe and the role of DE in the late time universe. In that model, the crossing of phantom divide occurs at the moment when the cosmological evolution of the dilaton yields the change of the sign of the total measure in the dilaton kinetic term of the underlying action, i.e. the phantom scalar field cosmology emerges in the course of dynamical evolution instead of introducing the wrong sign kinetic term into the action, as usual. }. We should note also that in this paper we restrict ourself with studying massive fermions without taking into account the Higgs mechanism. Preliminary study shows that in TMT,  in the context of cosmology, a possible   back reaction of fermions on the vacuum expectation value of the Higgs field may be relevant. In this respect the  model  of the present paper may be regarded as a toy one. It would be interesting to investigate in the future what kind of effect  such a back reaction may have on the change of the sign of the neutrino mass and on the possibility of the phantom-like scenario.

\textit{IV. The present value of the vacuum energy density}. The effect of the dynamical DE  was realized here on the basis of the first principles of TMT. The only additional quantitative assumption is that the dimensionless parameters $b$, $k$ and $h$ have the same order of magnitude. As we mentioned in the step I of Sec.IV, the observed tiny value of $\rho_{vac}$ may be achieved if these parameters are huge numbers. We want to argue here that the latter choice is very much different from what is usually called a fine tuning in resolution of the old or new CC problem\cite{Weinberg1}. When the CC problem  one hands over to particle field theory, the generally accepted understanding of the required subtraction or cancellation mechanism consists in a fine tuning of {\it coupling constants and masses in   similar  terms in the Lagrangian}, like e.g. as it is done in SUSY and SUGRA. The way large dimensionless numbers we would like to appear in the TMT action is absolutely different: 1) the parameters $b$, $k$ and $h$ are huge numbers as compared with dimensionless parameters in particle field theories but there is no need in mutual fine tuning of  $b$, $k$ and $h$; 2) huge values of
$b$, $k$ and $h$ have no any observable consequences except for the tiny value of the vacuum energy density in  comparison with the typical local matter energy density.

\textit{V. Generalization to the Standard Model matter content}.  In our earlier papers\cite{GK2-1} we have shown that the gauge fields may be added to the model without changing the constraint. The nonabelian symmetries are also compatible with TMT. Therefore the matter content and symmetries of the model may be extended up to those of the Standard Model.  We would like to stress again that results of the present paper allow us to claim that the effect of the dynamical DE  may be realized with the only matter content of the Standard Model and without any additional dynamical degrees of freedom intended to get this effect. Note that  all the peculiar TMT effects (new with respect to the conventional theory) emerging in the presence of fermions become unobservable in the regular particle physics conditions.

  \textit{VI. Fermion dark matter candidate?} One can finally propose the hypothesis concerning the nature of the dark matter. Creation of fermions in regular particle physics conditions implies that the fermion emerges in the "high density" regime. As we have mentioned in the item I of this section, in such a case the fermion may be in the state with $\zeta$ very close either to $\zeta_1$ or to  $\zeta_2$. In our earlier papers\cite{GK2-1} concerning studying the Standard Model in the framework of TMT, we explored the idea that fermion states with $\zeta$ very close to $\zeta_1$ and  $\zeta_2$ might be associated with the first and second fermion generations respectively. However in the present paper we run into a drastic difference in features of the $\zeta_1$ and  $\zeta_2$ fermion states.
 In fact, we have elucidated that if fermions have been created in the state with $\zeta$ very close to $\zeta_1$ and  satisfying the conditions (\ref{interval_zeta_A}) or (\ref{interval_zeta_B}) then they are able to evolve to the arbitrary low energy density. If  however fermions have been created in the state with $\zeta$ very close to $\zeta_2$ then it follows from Eq.(\ref{sign_bigger_zero}) that  only in two following cases it is possible: (1) $\zeta_0<\zeta_2<\zeta<\zeta_1$ and (2)  $-b<\zeta<\zeta_2<\zeta_0<\zeta_1$ \footnote{ The case (2) involves an additional requirement on the parameter $b$ which is satisfied in the numerical solutions of Sec.VI for the scenarios Ib, II and III.}.
 One can check using Eq.(\ref{diff-eq-zeta}) that $\zeta$ is forced to oscillate in the interval $\zeta_2<\zeta<\zeta_1$ in the case (1) and in the interval $-b<\zeta<\zeta_2$ in the case (2). Such regimes should perhaps be ended at some intermediate equilibrium values, but for us it is important here that the fermion starting from such a state is unable to evolve to a state with arbitrary low energy density. We come to this conclusion in the FLRW metric but one can expect a similar result e.g  in a spherically symmetric gravity problem.    Since $\zeta$ is supposed to be a continuous function, one can expect that there should be a mechanism which prevents a possibility  for the fermion in the states of the cases (1) and (2) to be in the same space-time region as well as for each of them to be in the same space-time region with another fermion whose state starts from  $\zeta$ satisfying the conditions (\ref{interval_zeta_A}) or (\ref{interval_zeta_B}). Suppose that the dark matter consists of the same fermions (from the viewpoint of the quantum field theory) as the visible fermion matter. Our hypothesis may be formulated in the following form:   the fermion in the state  with $\zeta$ satisfying the conditions (\ref{interval_zeta_A}) or (\ref{interval_zeta_B}) is the visible fermion while the same fermion but in the state as in  the cases (1) or (2) is a candidate for the dark matter particle. It may be that inability to exists in the same space-time region explains the absence of the visible to dark matter local interactions. Besides, if for example in the case (2), the equilibrium value that $\zeta$ reaches in the above mentioned oscillations  is close to $-b$ then the  mass of the  fermion  in such a state may be much bigger than the one of the visible fermion state (see Eq.(\ref{rhofcan})). Then physics of such fermions may be very much different from the regular particle physics. It would be interesting   to investigate a possibility to realize  conditions for an existence of a heavy sterile neutrino.

\section{Acknowledgments}
We thank J. Bekenstein, D. Polarski, S. Tsujikawa, J.-P. Uzan and A. Vilenkin for helpful conversation on the subjects of the paper.

\clearpage

\begin{figure}[htb]
\begin{center}
\includegraphics[width=8.0cm,height=6.0cm]{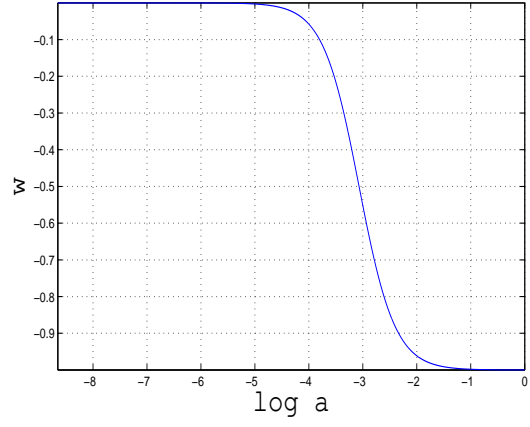}
\end{center}
\caption{\textit{In figures 1-4 the I.A type of scenario} with parameters: $\Lambda_0 =-1.5$, $k=2$, $h=0.3$,
 $M_{cl}^4=10^{-5}$. Then $\zeta_1=2.3114$, $\zeta_2 =-1.2114$. The initial value of $\bar{\zeta}$ is chosen
to be $\bar{\zeta}_{in}=2.312$. In this figure: the total EoS $w=\bar{p}/\bar{\rho}$  vs. $\ln a$.}
\end{figure}

\begin{figure}[htb]
\begin{center}
\includegraphics[width=12.0cm,height=6.0cm]{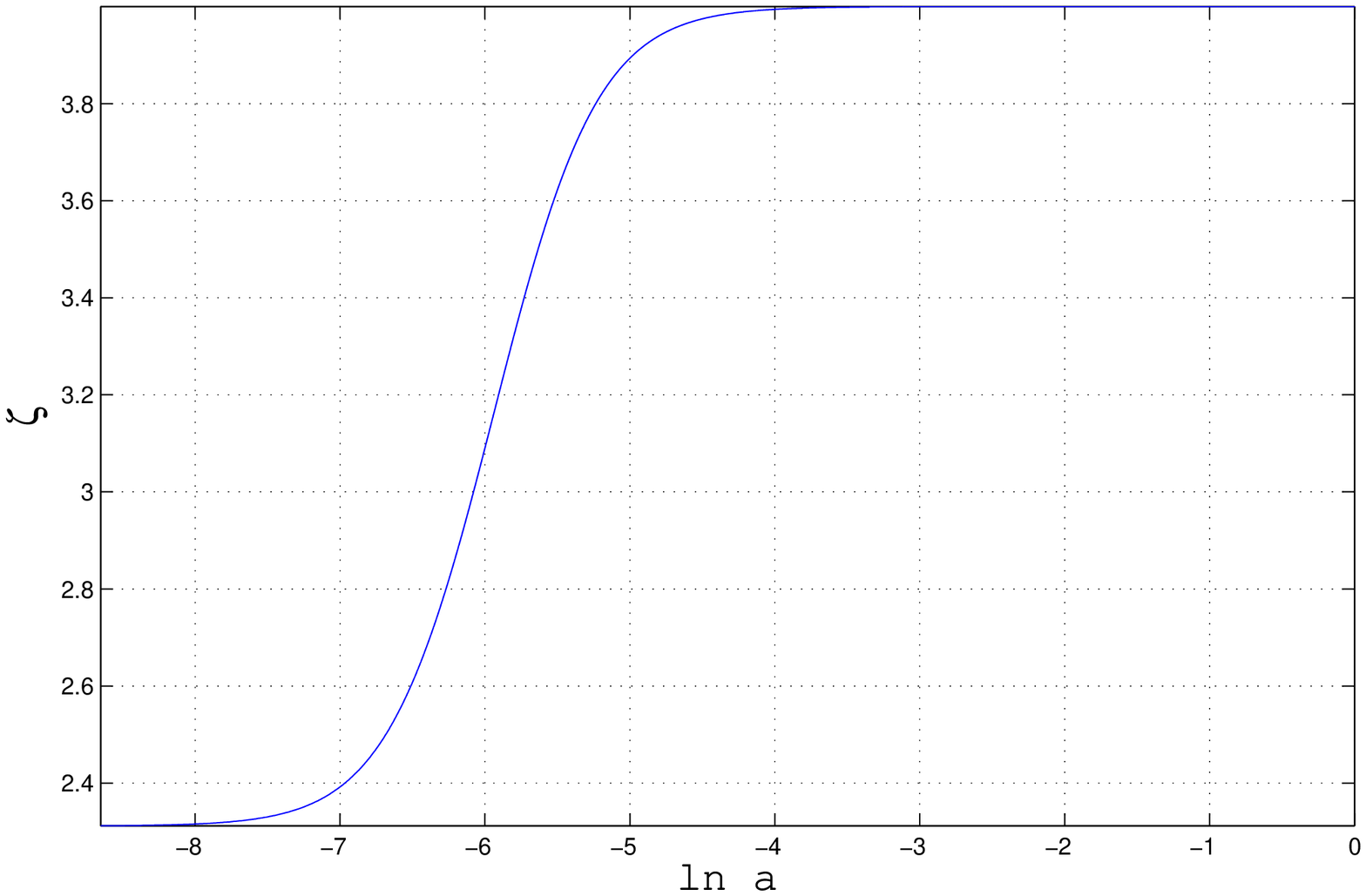}
\end{center}
\caption{ The I.A type of scenario. Evolution of the averaged value  $\bar{\zeta}$ that with very high accuracy coincides with the value of $\zeta$ in the maximal volume domains of low fermion density. The result of this numerical solution confirms the analytic estimations made in the end of Sec.IV and formulated under the name the "self-locking retainer" effect: as the total energy density decays from the value $\sim 10^{14}$ at the cold matter dominated epoch with $w\approx 0$ up to the value $\sim 10^{-2}$ at the DE dominated epoch with $w\approx -1$ (see Fig.3), $\bar{\zeta}$ changes only from $\bar{\zeta}\approx 2.3$ to the fermion vacuum value $\bar{\zeta}=\zeta_0 =4$ defined by Eq.(\ref{zete-vacuum}).}
\label{fig13}
\end{figure}

\label{fig1}
\begin{figure}[htb]
\begin{center}
\includegraphics[width=8.0cm,height=6.0cm]{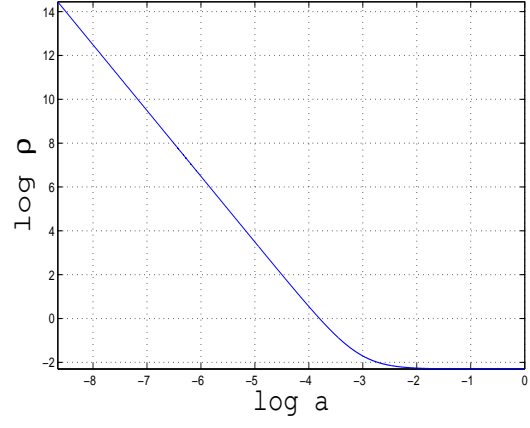}
\end{center}
\caption{The I.A type of scenario. Total energy density.}
\label{fig2}
\end{figure}
\begin{figure}[htb]
\begin{center}
\includegraphics[width=8.0cm,height=6.0cm]{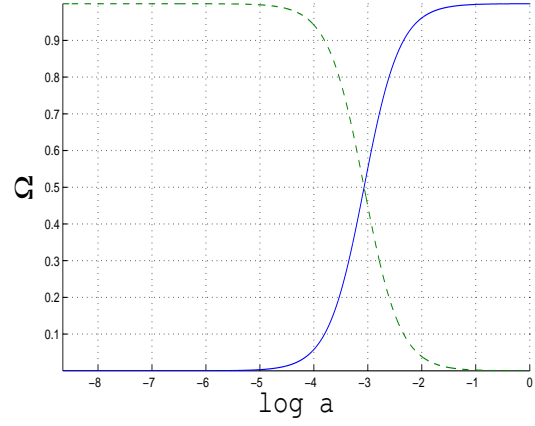}
\end{center}
\caption{The I.A type of scenario. $\Omega$ vs $\ln a$ where fractions of clustered (dark) matter $\Omega_m$ (the black dash line) and effective DE $\Omega_{DE}$ (the blue solid line).}
\label{fig14}
\end{figure}

\clearpage

\begin{figure}[htb]
\begin{center}
\includegraphics[width=8.0cm,height=6.0cm]{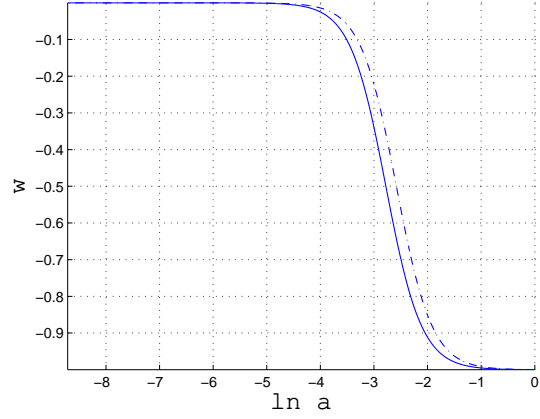}
\end{center}
\caption{\textit{In figures 5-10 the I.B type of scenario} with parameters: $\Lambda_0=-0.2$; $k=-0.3$; $h=-1$; $M_{cl}^4=5\cdot 10^{-5}$; Then $\zeta_1=3.0595$; $\zeta_2=-0.3595$. The initial value of $\bar{\zeta}$ is chosen
to be $\bar{\zeta}_{in}=3.0594$. In this figure: total EoS (solid line).  The cosmological dynamics practically coincides with that of the $\Lambda$CDM model which is presented in this figure and in Figs. 6-8, 10 by dot-dashed lines. In order to avoid superimposition of the graphs of the TMT and $\Lambda$CDM models, the initial values of the cold (dark) matter energy density in $\Lambda$CDM model have been chosen to be slightly bigger than those in the appropriate scenario of our TMT model}
\label{fig8}
\end{figure}
\begin{figure}[htb]
\begin{center}
\includegraphics[width=12.0cm,height=6.0cm]{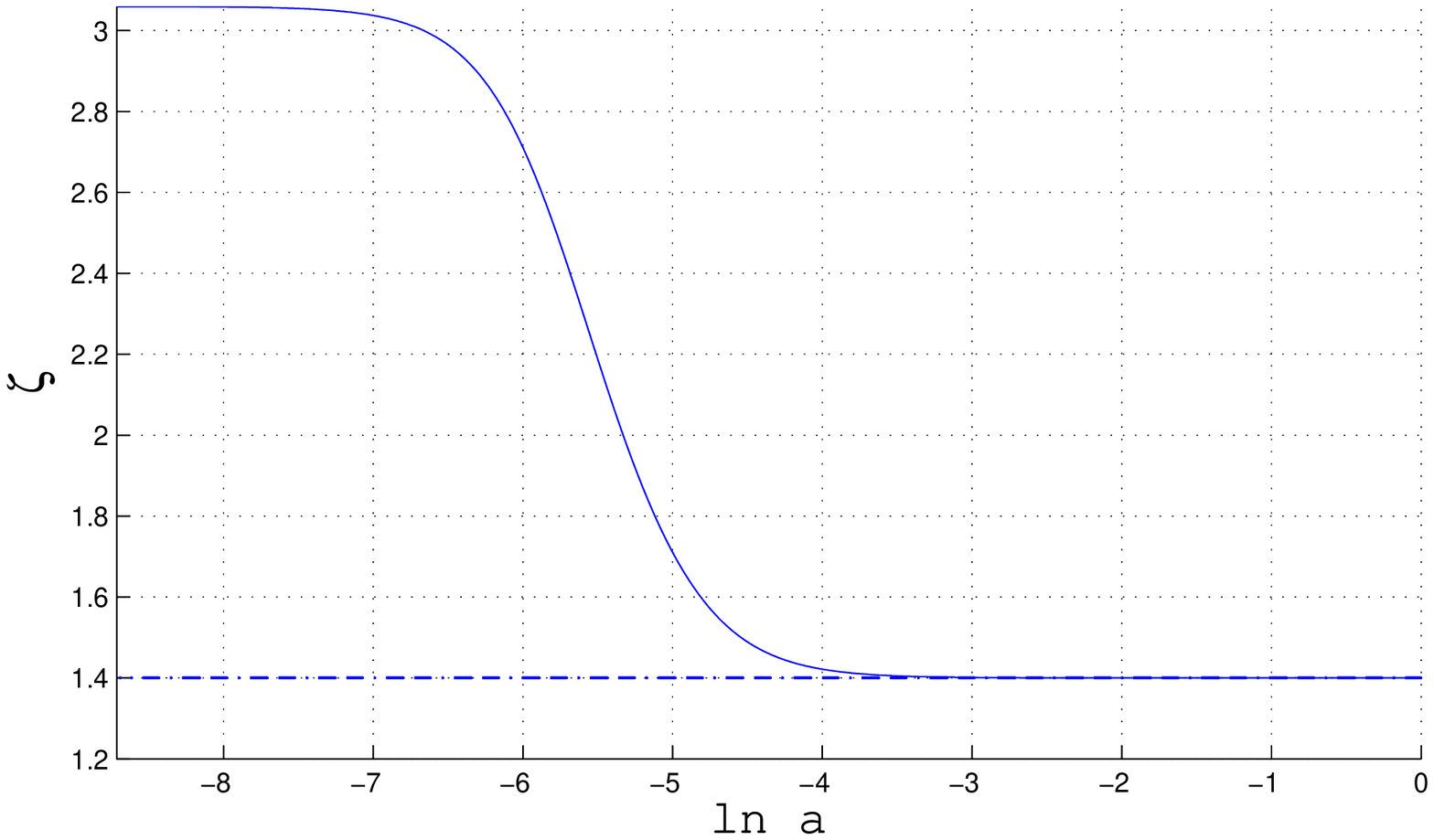}
\end{center}
\caption{  The I.B type of scenario.  Evolution of the averaged value  $\bar{\zeta}$ that with very high accuracy coincides with the value of $\zeta$ in the maximal volume domains of low fermion density. The result of this numerical solution confirms the analytic estimations made in the end of Sec.IV and formulated under the name the "self-locking retainer" effect: as the total energy density decays from the value $\sim 10^{16}$ at the cold matter dominated epoch with $w\approx 0$ up to the value $\sim 10^{-2}$ at the DE dominated epoch with $w\approx -1$ (see Fig.8), $\bar{\zeta}$ changes only from $\bar{\zeta}\approx 3.06$ to the fermion vacuum value $\bar{\zeta}=\zeta_0 =1.4$ defined by Eq.(\ref{zete-vacuum}). The dot-dashed line shows the constant $\bar{\zeta}=\zeta_0 =1.4$ which is used to reduce our TMT model to the appropriate $\Lambda$CDM model. }
\label{fig12}
\end{figure}
\begin{figure}[htb]
\begin{center}
\includegraphics[width=14.0cm,height=10.0cm]{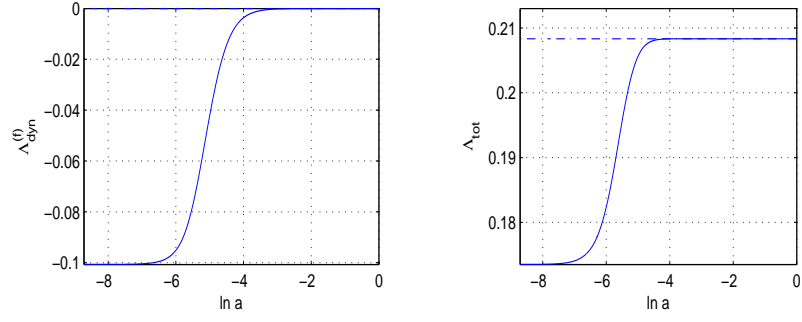}
\end{center}
\caption{The I.B type of scenario. The averaged values of the dynamical fermionic $\Lambda$ term $\overline{\Lambda_{dyn}^{(f)}}$, Eq.(\ref{Lambda-canon-av}), and of the total CC $\Lambda_{tot}$ vs.  $\ln a$. The result of this numerical solution confirms  the "self-locking retainer" effect. The dot-dashed line shows the constant CC used in the appropriate $\Lambda$CDM model.}
 \label{fig9}
\end{figure}


\begin{figure}[htb]
\begin{center}
\includegraphics[width=8.0cm,height=6.0cm]{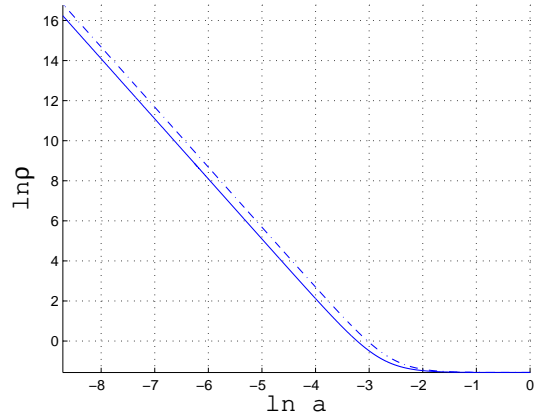}
\end{center}
\caption{ The I.B type of scenario. The total energy density.}
\label{fig10}
\end{figure}
\begin{figure}[htb]
\begin{center}
\includegraphics[width=14.0cm,height=10.0cm]{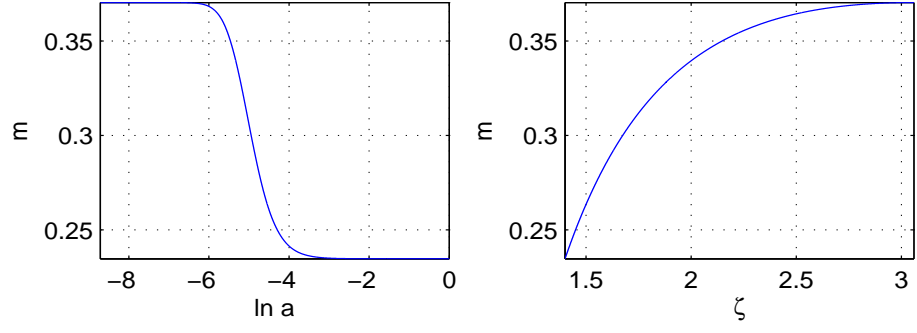}
\end{center}
\caption{ The I.B type of scenario.  Mass of the cold neutrinos in the maximal volume domains of low density: $m$ vs $\ln a$ and $m$ vs $\bar{\zeta}$.}
\label{fig11}
\end{figure}

\begin{figure}[htb]
\begin{center}
\includegraphics[width=8.0cm,height=6.0cm]{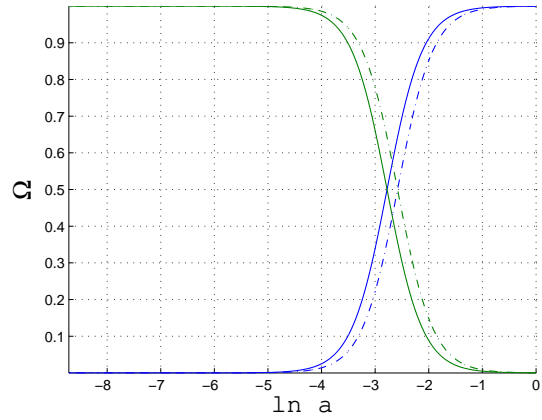}
\end{center}
\caption{The I.B type of scenario.  $\Omega$ vs $\ln a$ where fractions of clustered (dark) matter $\Omega_m$ (the red solid line) and effective DE $\Omega_{DE}$ (the blue solid line).   The dot-dashed lines show $\Omega_m$ and $\Omega_{DE}$ for the appropriate $\Lambda$CDM model. }
\label{fig13}
\end{figure}

\clearpage

\begin{figure}[htb]
\begin{center}
\includegraphics[width=8.0cm,height=6.0cm]{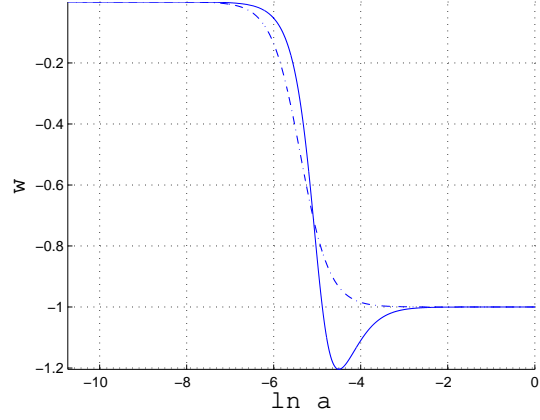}
\end{center}
\caption{\textit{In figures 11-16 the II, phantom-like type of scenario} with parameters: $\Lambda_0=-0.2$; $k=-0.3$; $h=-6$; $M_{cl}^4=5\cdot 10^{-8}$; Then $\zeta_1=18.2267$; $\zeta_2=-0.5267$. The initial value of $\bar{\zeta}$ is chosen
to be $\bar{\zeta}_{in}=18.2266$. In this figure: the total EoS (solid line) typical for \textit{the phantom-like scenario} of the late universe. In this figure and in Figs. 12-14, 16 the results for the appropriate $\Lambda$CDM model are presented  by dot-dashed lines.}
 \label{fig14}
\end{figure}
\begin{figure}[htb]
\begin{center}
\includegraphics[width=10.0cm,height=6.0cm]{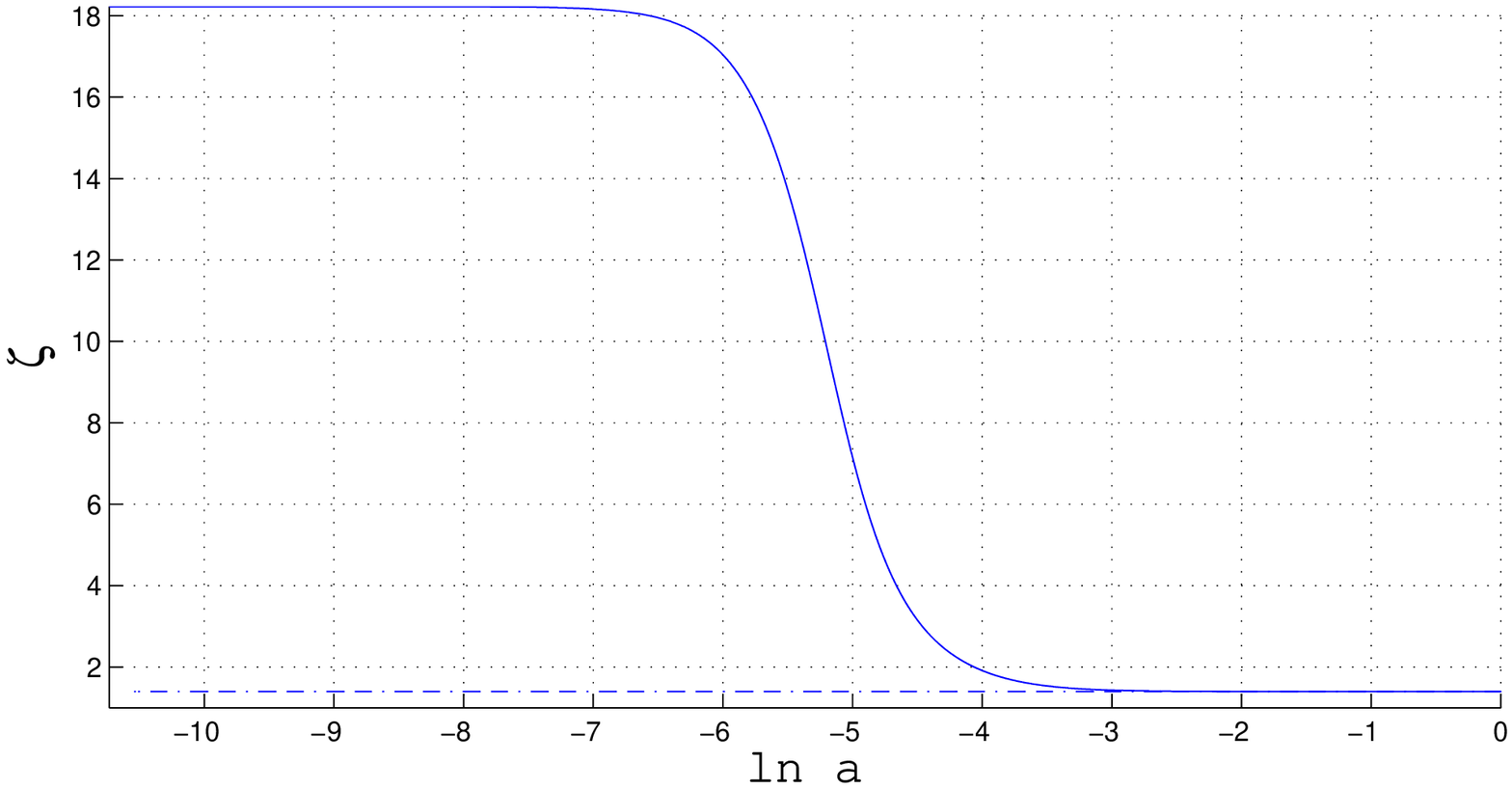}
\end{center}
\caption{ The II, phantom-like type of scenario.   Evolution of the averaged value  $\bar{\zeta}$ that with very high accuracy coincides with the value of $\zeta$ in the maximal volume domains of low fermion density. The result of this numerical solution confirms the analytic estimations made in the end of Sec.IV and formulated under the name the "self-locking retainer" effect: as the total energy density decays from the value $\sim 10^{11}$ at the cold matter dominated epoch up to the value $\sim 10^{-2}$ at the DE dominated epoch (see Fig.14), $\bar{\zeta}$ changes only from $\bar{\zeta}\approx 18$ to the fermion vacuum value $\bar{\zeta}=\zeta_0 =1.4$ defined by Eq.(\ref{zete-vacuum}).}
\label{fig19}
\end{figure}

\begin{figure}[htb]
\begin{center}
\includegraphics[width=14.0cm,height=10.0cm]{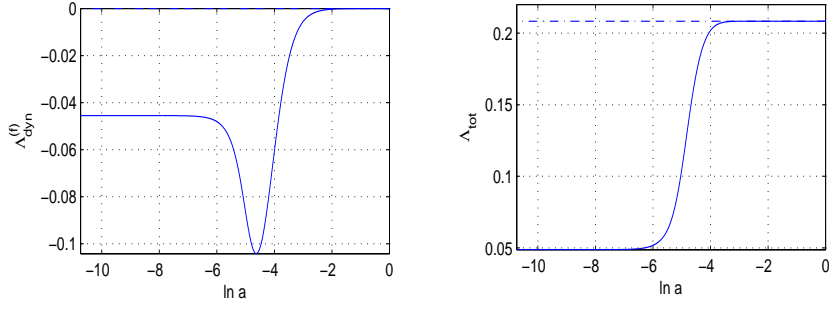}
\end{center}
\caption{The II, phantom-like type of scenario.  The averaged values of the dynamical fermionic $\Lambda$ term $\overline{\Lambda_{dyn}^{(f)}}$, Eq.(\ref{Lambda-canon-av}), and of the total CC $\Lambda_{tot}$ vs.  $\ln a$. The result of this numerical solution confirms  the "self-locking retainer" effect.}
\label{fig16}
\end{figure}
\begin{figure}[htb]
\begin{center}
\includegraphics[width=8.0cm,height=6.0cm]{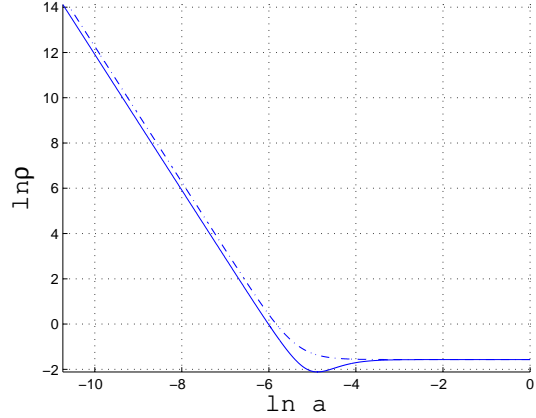}
\end{center}
\caption{ The II, phantom-like type of scenario.  The result of the numerical solution for the total average energy density $\bar{\rho}$ confirms the analytic estimations made in Sec.V: $\bar{\rho}$  passes through minimum that is followed by monotonic increasing with $\bar{\rho}\rightarrow\rho_{vac}$ from below. Such a  behavior of $\bar{\rho}$  is typical  for \textit{the phantom like scenario} where the very late universe evolves as in \textit{the pseudo-rip model}\cite{Frampton}.}
\label{fig17}
\end{figure}
\begin{figure}[htb]
\begin{center}
\includegraphics[width=18.0cm,height=10.0cm]{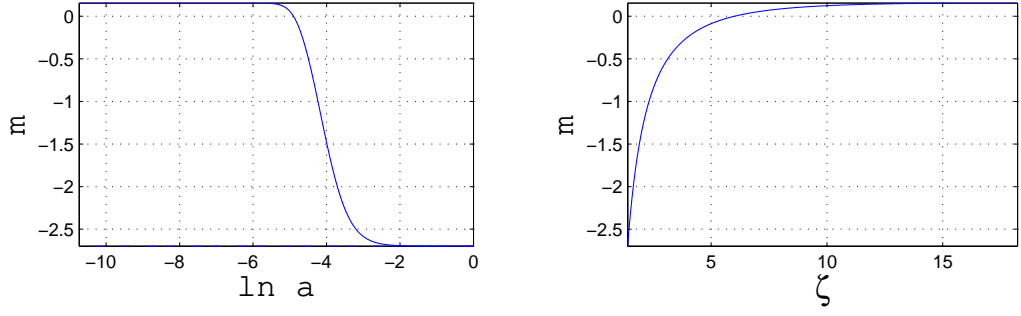}
\end{center}
\caption{  The II, phantom-like type of scenario.  Mass of the cold neutrinos in the maximal volume domains of low density: $m$ vs $\ln a$ and $m$ vs $\bar{\zeta}$. It is seen that $m$ changes sign and asymptotically approaches a finite negative value.}
\label{fig18}
\end{figure}

\begin{figure}[htb]
\begin{center}
\includegraphics[width=8.0cm,height=6.0cm]{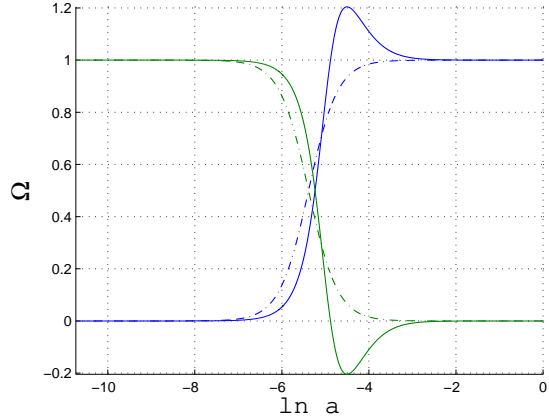}
\end{center}
\caption{The II, phantom-like type of scenario.   $\Omega$ vs $\ln a$ where fractions of clustered (dark) matter $\Omega_m$ (the red solid line) and effective DE $\Omega_{DE}$ (the blue solid line).  The simultaneous effects of exceedance of the value 1 for $\Omega_{DE}$ and lowering below the value $-1$ for $\Omega_m$ result from changing sign of the neutrinos mass and their canonical energy density in the maximal volume domains of low fermion density. The dot-dashed lines show $\Omega_m$ and $\Omega_{DE}$ for the appropriate $\Lambda$CDM model.}
\label{fig20}
\end{figure}

\begin{figure}[htb]
\begin{center}
\includegraphics[width=14.0cm,height=10.0cm]{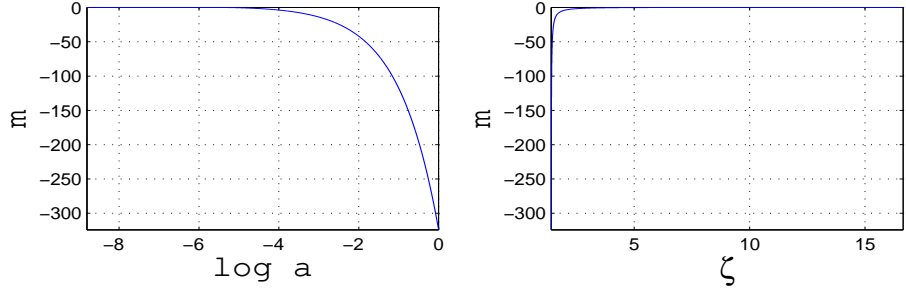}
\end{center}
\caption{ \textit{The III, phantom-like type of scenario of the cosmological evolution} with the fine tuned parameters such that $\zeta_0 =-k$: $\Lambda_0=-0.2$; $k=-1.4$; $h=-6$; $M_{cl}^4=5\cdot 10^{-8}$; Then $\zeta_1=16.6481$; $\zeta_2=-0.0481$. The initial value of $\bar{\zeta}$ is chosen
to be $\bar{\zeta}_{in}=16.647$.   In this figure we show the evolution of the mass of the cold neutrinos in the maximal volume domains of low density: $m$ vs $\ln a$ and $m$ vs $\bar{\zeta}$. $m$ changes sign from positive to negative and  $m\rightarrow -\infty$. Cosmological evolution of all other quantities is similar to that of the II, phantom-like type of scenario presented in Figs.11-16.}
\label{fig21}
\end{figure}

\end{document}